\documentclass{amsart}
\usepackage{amssymb}
\usepackage{amsfonts}
\usepackage{amsmath}
\usepackage[legalpaper,bookmarks=true,colorlinks=true,linkcolor=blue,citecolor=blue]{hyperref}
\usepackage{graphicx}%
\usepackage{fancyhdr}
\usepackage{color}
\usepackage[mathlines]{lineno}
\usepackage{lscape}
\usepackage{epsfig}
\usepackage{natbib}
\usepackage{geometry}
\usepackage{tgbonum}
\usepackage[]{listings}
\usepackage{enumitem}
\usepackage{multirow}

\fontfamily{qcr}\selectfont

\newtheorem{theorem}{Theorem}
\theoremstyle{plain}

\newtheorem{proposition}{Proposition}

\numberwithin{equation}{section}
\newtheorem{assumption}{Assumption}

\newcommand{\Bin}{\bigskip \noindent}

\newcommand{\Ni}{\noindent}

\begin{document}
\Large
\title[Two samples problems under Heavy-Tailed data]{Asymptotic Theory and Statistical Inference for the Samples Problems with Heavy-Tailed Data using the Functional Empirical Process}
\normalsize
\author{Abdoulaye CAMARA $^{(1)}$}
\author{Saliou DIOUF $^{(2)}$}
\author{Moumouni DIALLO $^{(3)}$}
\author{Gane Samb Lo $^{(4)}$}

\maketitle
\Ni\textbf{Abstract}. This paper introduces the Trimmed Functional Empirical Process (TFEP) as a robust framework for statistical inference when dealing with heavy-tailed or skewed distributions, where classical moments such as the mean or variance may be infinite or undefined. Standard approaches including the classical Functional Empirical Process (FEP), break down under such conditions, especially for distributions like Pareto, Cauchy, low degree of freedom Student- t, due to their reliance on finite-variance assumptions to guarantee asymptotic convergence. \\

\Ni The TFEP approach addresses these limitations by trimming a controlled proportion of extreme order statistics, thereby stabilizing the empirical process and restoring asymptotic Gaussian behavior. We establish the weak convergence of the TFEP under mild regularity conditions and derive new asymptotic distributions for one-sample and two-sample problems. These theoretical developments lead to robust confidence intervals for truncated means, variances, and their differences or ratios.\\

\Ni The efficiency and reliability of the TFEP are supported by extensive Monte Carlo experiments and an empirical application to Senegalese income data. In all scenarios, the TFEP provides accurate inference where both Gaussian-based methods and the classical FEP break down. The methodology thus offers a powerful and flexible tool for statistical analysis in heavy-tailed and non-standard environments.\\

\noindent\textbf{Keywords}. asymptotic theory; robust statistics; heavy-tailed distributions; functional empirical process; two-sample problems. \\
\noindent \textbf{AMS 2010 Mathematics Subject Classification:} 
Primary: 60F17, 62G32; Secondary: 62G10, 62F05, 60G50.

\newpage
\Ni $^{(1)}$ \textit{Abdoulaye Camara}, M.Sc., is preparing a Ph.D thesis under the supervision of the third author. He is a lecturer at Universit\'e des Sciences, Techniques et des Technologies, Bamako. He is also a junior researcher at: IMHOTEP International Mathematical Centre (IMHO-IMC: imhotepsciences.org)\\
Email: camara.abdoulaye2@ugb.edu.sn\\

\noindent $^{(2)}$ Pr Saliou DIOUF\\
Lecturer-researcher at the UFR SEFS\\
Gaston Berger University, Saint-Louis, S\'en\'egal\\
Email: saliou.diouf@ugb.edu.sn\\

\noindent $^{(3)}$ Dr Moumouni Diallo\\
Universit\'e des Sciences Sociale et de Gestion de Bamako ( USSGB)\\
Facult\'e des Sciences \'Economiques et de Gestion (FSEG)\\
IMHOTEP International Mathematical Centre (IMHO-IMC: imhotepsciences.org)\\
Email: moudiallo1@gmail.com.\\

\noindent $^{\dag}$ Pr Gane Samb Lo.\\
LERSTAD, Gaston Berger University, Saint-Louis, S\'en\'egal (main affiliation).\newline
LSTA, Pierre and Marie Curie University, Paris VI, France.\newline
AUST - African University of Sciences and Technology, Abuja, Nigeria\\
IMHOTEP International Mathematical Centre (IMHO-IMC: imhotepsciences.org)\\
gane-samb.lo@edu.ugb.sn, gslo@aust.edu.ng, ganesamblo@ganesamblo.net\\
Permanent address : 1178 Evanston Dr NW T3P 0J9, Calgary, Alberta, Canada.\\

\newpage
\Ni\textbf{Presentation of authors}.\\

\noindent \textbf{Abdoulaye Camara}, M.Sc., is an assistant professor of Mathematics and Statistics at: Universit\'e des Sciences, Techniques et des Technologies, Bamako. He is preparing a PhD thesis at: LERSTAD, Unversit\'e Gaston Berger Saint-louis, Senegal. He is also a junior researcher at: Mathematical Center (imho-imc), https://imhotepsciences.org.\\

\noindent \textbf{Pr Saliou DIOUF}, PhD, is a full Professor of mathematics and statistics at University Gaston Berger, Saint-Louis, Senegal.\\

\noindent \textbf{Moumouni Diallo}, PhD, is a Professor of Mathematics and Statistics at:Universit\'e des Sciences Sociales et de Gestion, Bamako, MALI. He is a senior researcher at: Mathematical Center (imho-imc), https://imhotepsciences.org.\\

\Ni \textbf{Gane Samb Lo}, Ph.D., is a retired full professor from Université Gaston Berger (2023), Saint-Louis, SENEGAL. He is the founder and the Probability and Statistics Chair holder at: Imhotep International Mathematical Center (imho-imc), https://imhotepsciences.org.\\
\newpage

\section{Introduction} \label{sec_01}

\Ni The statistical theory of general linear models, particularly analysis of variance (ANOVA) and linear regression, relies primarily on the assumption of data normality. This framework of Gaussian linear models, extensively documented in the literature (e.g., \cite{Seber2003, Scheffe1959, Tibshirani2009, Bruce2008}), is both elegant and powerful. It provides a solid theoretical foundation for inference, ensuring, for example, the convergence of central limit theorems and the stability of the underlying stochastic processes. \\

\Ni However, in many real-world applications such as economics, insurance, finance, and environmental science, data rarely conform to the Gaussian paradigm. In that case, non-Gaussian data frequently generate with heavy tails, skewness, or mixtures of distributions. For example, accident data can often be modeled by Gamma law, log-normal law or Pareto law can be used for income. In these frameworks, classical Gaussian assumptions become unreliable implying unstable or biased results for standard inference tools. \\

\Ni As illustrate, we consider income data of Senegalese ESAM1 (1996) and ESAM2 (2000) for the Dakar and Diourbel regions. The corresponding datasets are denoted $\texttt{dakar1}$, $\texttt{dakar2}$, $\texttt{diour1}$, and $\texttt{diour2}$, respectively. In this study, we analyze a sample of $T$ observations from each dataset for illustrative purposes.

\Ni Our objective are: (i) to compare income distributions between the two regions and (ii) to examine the temporal evolution of income. As shown later, these datasets have non-Gaussian behavior, characterized by skewness, kurtosis, and heavy tails.\\

\Ni In such circumstances, Gaussian-based inferential methods become unreliable. The assumption of finite moments, particularly a finite variance. This is the central idea to classical asymptotic theory, often fails in heavy-tailed or contaminated contexts. Consequently, the empirical mean and variance may lose their convergence properties, leading to unstable estimators and invalid inference. Standard procedures such as the t-test or ANOVA may then yield inflated variances, distorted coverage probabilities, and even inconsistent results.\\

\Ni Formally, the classical Functional Empirical Process (FEP) is defined as

\[
\mathbb{G}_n(f) = \sqrt{n}(\mathbb{P}_n - \mathbb{P})(f),
\]

\Ni Under conditions of regularity, the process $\mathbb{G}_n$ converges to a tight Gaussian limit \cite{VaartWellner1996}. This justify the use of classical functional empirical process. However, when the underlying distribution have infinite variance, the process $\mathbb{G}_n$ do not converge, leading to the breakdown of the FEP framework.\\

\Ni To address such issues, robust asymptotic methods based on weak convergence theory have been developed (e.g., \cite{Lo2018FEP, loFepLrepMethod, AbdoulayeFEP2023}). These methods provide powerful probabilistic tools for analyzing the limiting distributions of statistics and for constructing consistent estimators and valid hypothesis tests even under non-standard conditions. Central to this framework are the Empirical Distribution Function (EDF), the Empirical Process (EP), and its generalization, the Functional Empirical Process (FEP) \cite{VaartWellner1996, VaartWellner2023, Lo2016}. These provide a unified language for studying the asymptotic behavior of estimators and test statistics in general settings.\\

\Ni Nevertheless, the classical FEP approach remains inadequate when dealing with infinite-variance or strongly skewed data, such as those following Pareto, Cauchy, or low-degree Student-$t$ distributions. In such contexts, the regularity conditions required for weak convergence are violated, leading to a loss of tightness and divergence of empirical functionals.\\

\Ni To overcome these limitations, we propose a robust alternative, the Trimmed approach (e.g., \cite{Dhar2009, Dhar2022, Oliveira2025, Michael2017, Mason1989}). The principe is to apply trimming before statistical processing, thereby stabilizing the empirical process while preserving the essential structure of the data.
This trimming ensures that the process retains its tightness and satisfies central functional limit theorems, even when the underlying distribution exhibits heavy tails or infinite variance. The resulting TFEP framework extends the classical functional empirical process machinery to a broader class of distributions and provides a solid foundation for robust inference in non-standard environments.\\

\Ni The remainder of this article is organized as follows.  Section \ref{sec_02} introduces the fundamental difficulties posed by heavy-tailed distributions and explains why classical asymptotic theory fails in this setting. Section \ref{sec_03} develops the trimmed Functional Empirical Process (TFEP), establishing its main properties and providing the necessary theoretical tools for robust inference under infinite-variance conditions. In Section \ref{sec_04}, we derive asymptotic estimators for key parameters including the mean, variance, and tail-sensitive functionals together (difference of means and variances ratio) with their non-Gaussian limit distributions. Section~\ref{sec_05} reports simulation studies an empirical application that evaluate the finite-sample performance of the proposed methods across a variety of heavy-tailed and skewed  models. Finally, Section~\ref{sec_06} concludes with a discussion of practical implications, limitations, and directions for future research. \\

\section{Background and Preliminaries} \label{sec_02}

\subsection{The Core Problem: Why Heavy Tails Break Classical Theory}  

\Ni \\

\Ni Standard asymptotic statistical theory for statistical inference relies on the Central Limit Theorem (CLT), which requires finite variance. For some heavy-tailed distributions such that Pareto with exponent $\alpha \le 2$, Cauchy, L\'evy, Student-t with low df, we often have: 

\[
\mathbb{E}[X^2] = \int x^2 dF(x) = \infty, \quad \text{or even } \quad  \mathbb{E}[|X|] = \infty  
\]

\Bin This means the variance is infinite and the classical CLT fails. In that case, the sample mean  becomes instable and does not converges at rate $\sqrt{n}$. In some the sample mean converges at rate $n^{1/\alpha}$, when $1< \alpha < 2$(Pareto law) or does not converge at all (Cauchy case).\\

\Ni Thus the classical normalization $\sqrt{n}(\overline{X}_n) - \mu)$ does not converge to a Gaussian limit. 

\Ni So CLT Limit Changes and Stable Laws can be used instead of the Gaussian law and we enter the domain of the Generalized Central Limit Theorem. It states that if $F$ is in the domain of attraction of a stable law with index $\alpha \in (0,2]$, then there exist sequences $a_n > 0$ and $b_n$ such that:

\[
a_n^{-1} \left(\sum_{i=1}^n X_i - b_n\right) \xrightarrow{d} S_{\alpha}
\]

\Bin where $S_{\alpha}$ is an $\alpha$-stable distribution, not normal (e.g., \cite{Gennady1994, Nolan2020, Resnick2007}).\\

\Ni This non-standard limit theory complicates statistical inference for common parameters like the mean, quantiles, and regression coefficients.\\

\subsection{Heavy-tailed distributions}

\Ni We now recall some notions about the behavior of heavy-tailed data. In modern probability and statistics heavy-tailed distributions play a central role, especially in finance, insurance, environmental data, and income modeling. Their defining feature is the slow decay of the tail probability. In this part, we introduce some key concepts required for Functional Empirical Process (FEP) theory under heavy tails.\\

\Ni The tails of a distribution are the parts far away from the mean (center) of the distribution: \\

\begin{itemize}
	\item \textbf{Left tail}: Extremely small values.\\
	
	\item \textbf{Right tail}: Extremely large values. \\
\end{itemize}

\Ni Formally, for a random variable $X$ with cumulative distribution function (CDF) $F(x)$

\begin{itemize}
	\item \textbf{Left tail}: $\mathbb{P}(X < x) = F(x),$\\
	
	\item \textbf{Right tail}: $\mathbb{P}(X > x) = 1 - F(x)$.
\end{itemize}

\Bin\textbf{Definition and Basic Properties}\\

\Ni A distribution $F$ on $\mathbb{R}$ is called heavy-tailed if its right tail $\mathbb{P}(X > x) = 1 - F(x)$ decays more slowly than any exponential function. Formally: \\

\Ni A distribution $F$ heavy-tailed if, for all $\lambda > 0 $, 

\[
\lim_{x \to \infty} e^{\lambda x} F(x) = \infty. 
\]

\Bin This class includes power-law, Pareto-type law, and regularly varying distributions. They are characterized by high probability of extreme values and often infinite moments.\\

\Ni A leading subclass of heavy-tailed distributions is the family of regularly varying tails.\\

\Ni A distribution $F$ is said to have a regularly varying tail with index $\alpha > 0$ if 

\[
F(x) \sim L(x) x^{- \alpha} \quad \text{as} \quad x \to \infty,
\]

\Ni where

\begin{itemize}
	\item $\alpha$ is the tail index.\\
	
	\item $L(x)$ is a slowly varying function 
	
	\[
	\lim_{x \to \infty} \frac{L(tx)}{L(x)} = 1 \quad \text{for any} \quad t > 0
	\]
\end{itemize}

\Ni The parameter $\alpha$ measures tail heaviness and its magnitude determines the existence of moments. Small $\alpha \le 1$ implies heavier tail (more extreme events), that menas no finite mean. For $1 < \alpha \le 2$, there are finite mean, but infinite variance and classical CLT works for $\alpha > 2$ (finite mean and variance). \\

\Ni Under heavy tails, classical empirical process theory fails in several ways: 

\begin{itemize}
	\item Classical Donsker theorem may break down.\\
	
	\item Extreme values dominate.\\
	
	\item Limiting laws mix Gaussian and stable components.
\end{itemize}

\Bin To restore asymptotic normality, we truncate large and / or small observations. 
Thus arises the Truncated Functional Empirical Process (TFEP). In our study, we focus on the trimming approach,, which is a truncation version. \\

\Ni We know that the non-Gaussian behavior in the heavy tails data may be caused by some large observations (extremes) controlling or modeling them yields tractable limits. In our study, we focus on the trimming approach for heavy-tails data in order to make inference on the mean and variance. Next is the trimmed functional empirical process. \\

\section{Trimmed Functional Empirical Process}\label{sec_03}

\subsection{Preliminaries}

\Ni \\

\Ni Let $(\Omega, \mathcal{F}, \mathbb{P})$ be a probability space, and let $X, X_1, X_2, \dots$ be independent and identically distributed random variables from  distribution $F$ and taking values in a measurable space $(\mathcal{X}, \mathcal{A})$. We consider a class of measurable functions $\mathcal{F} = \{f: \mathcal{X} \to \mathbb{R}\}$ that will index our empirical process.\\

\Ni The empirical measure $\mathbb{P}_n$ is defined as:
\begin{equation}
	\mathbb{P}_n(f) = \frac{1}{n} \sum_{i=1}^n f(X_i).
\end{equation}

\Ni It is a random probability measure on $(\mathcal{X}, \mathcal{A})$, satisfying: 

\[
\mathbb{P}_A = \frac{1}{n} \sum_{i=1}^n 1_A(X_i), \quad A \in \mathcal{A}.
\]

\Ni The functional empirical process (FEP) is defined as:

\begin{equation}
	\mathbb{G}_n(f) = \sqrt{n} (\mathbb{P}_n - \mathbb{P})(f) = \frac{1}{\sqrt{n}} \sum_{i=1}^n \left(f(X_i) - \mathbb{E}[f(X_i)]\right), \quad f \in \mathcal{F}
\end{equation}

\Ni where $\mathbb{P}(f) = \mathbb{E}[f(X)]$ and $\mathbb{P}_n(f) = \frac{1}{n} \sum_{i=1}^n f(X_i)$. \\

\Ni Under regularity conditions (e.g., boundedness, finite bracketing entropy), $\mathbb{G}_n$ converges weakly in $\ell^{\infty}(\mathcal{F})$ to a tight, mean-zero Gaussian process $\mathbb{G}$ with covariance 

\[
Cov(\mathbb{G}(f), \mathbb{G}(g)) = Cov(f(X), g(X)).
\]

\Bin However, this classical process is sensitive to outliers or infinite variance since large values of $f(X_i)$ can dominate the sum.

\Ni The standard FEP theory requires finite second moments and light-tailed distributions to ensure tightness and convergence. When dealing with heavy-tailed data, these conditions are violated and the process $\mathbb{G}_n$ does not converge in distribution. In such cases, the standard empirical process becomes unreliable. Consider the following examples: \\

\Ni\textbf{Example}[Pareto Distribution]
Let $X \sim \text{Pareto}(\alpha, x_m)$ with survival function $\mathbb{P}(X > x) = (x_m/x)^\alpha$ for $x \geq x_m$. For $\alpha \leq 2$, the variance is infinite, and the classical FEP fails to converge to a tight Gaussian process. \\

\Ni\textbf{Example}[Student's t-distribution]
For $X \sim t_\nu$ with $\nu \leq 2$, the variance is infinite, and the standard empirical process exhibits erratic behavior.\\

\Ni To address these issues, we can introduce a robust mechanism that stabilizes the empirical process while preserving its essential structure. There exists different robustification strategies, such as trimmed approach, Winsorization method, M-estimators, Scale-robust methods, etc. Here we treat the trimming approach, which removes the extreme values in a dataset before conducting estimation or inference.\\

\subsection{General Framework}

\Ni Consider an i.i.d. random sample \( X_1, \dots, X_n \) defined on a measurable space $(\mathcal X, \mathcal A)$, with distribution $F$ on $\mathbb{R}$, having a density function $f$ that is continuous and strictly positive in a neighborhood of the $\tau$ - and $(1 - \tau)$ - quantiles for fixed $\tau \in [0, 1/2)$ a symmetric trimming proportion. Let $X_{(1)} \leq X_{(2)}, \dots, X_{(n)}$ represent the order statistics. We assume that a trimming threshold $k_n$ is fixed such that 

\[
k_n \to \infty, \quad \frac{k_n}{n} \to \tau,
\]

\Bin where $k_n = \lfloor \tau n \rfloor$.\\

\Ni Formaly, define the symmetric trimmed sample as 

\[
X_{(k_n +1)}, X_{(k_n +2)}, \ldots, X_{(n - k_n)}
\]

\Ni We keep only the central $n - 2k_n$ bservations.\\

\Ni The trimmed mean is defined as

\begin{equation*}
	\bar{X}_{n, \tau} = \frac{1}{n_\tau} \sum_{i=k_n +1}^{n-k_n} X_{(i)},
\end{equation*}

\Ni where $n_\tau = n - 2k_n$ is the effective size after trimming. \\

\Ni The corresponding population trimmed mean is: 

\[
\mu_\tau = \frac{1}{1-2 \tau} \int_{\tau}^{1-\tau} F^{-1}(u) du =  \frac{1}{1-2\tau} \int_{F^{-1}(\tau)}^{F^{-1}(1-\tau)} x \, dF(x) 
\]

\Bin The trimmed sample variance is defined as:

\[
S_{n, \tau}^2 = \frac{1}{n - 2k_n -1} \sum_{i=k_n +1}^{n-k_n} \left(X_{(i)} - \bar{X}_{n, \tau}\right)^2
\]

\Bin The corresponding population trimmed variance is: 

\[
\sigma_\tau^2
=
\frac{1}{1-2\tau} \int_{F^{-1}(\tau)}^{F^{-1}(1-\tau)} (x - \mu_\tau)^2 \, dF(x)
=
\frac{1}{1-2\tau} \int_{\tau}^{1-\tau} \big(F^{-1}(u) - \mu_\tau\big)^2 \, du
\]

\Bin The estimators of trimmed mean and trimmed variance use only the central part of the distribution (middle $1-2 \tau$ fraction), excluding the tails. \\

\subsection{Functional Empirical Process Framework} 

\Ni The symmetric trimming empirical measure is defined as

\begin{equation}
	\mathbb{P}_{n_\tau}(f) = \frac{1}{n_\tau} \sum_{i=k_n+1}^{n-k_n} f(X_{(i)}), \quad f \in \mathcal{F}.
\end{equation} 

\Ni This formulation corresponds to symmetric two-sided trimming; one-sided trimming schemes can be defined analogously. The trimming parameter $\tau$ governs the fundamental trade-off between robustness.\\

\Ni The Trimming Functional Empirical Process (TFEP) is defined by  

\begin{equation}
	\mathbb{G}_{n_\tau}(f) 
	= \sqrt{n_\tau} \left(\mathbb{P}_{n_\tau}(f) - \mathbb P_\tau(f) \right) = \frac{1}{\sqrt {n_\tau}} \sum_{i=k_n +1}^{n - k_n}\left(f(X_{(i)}) - \mathbb{E}[f(X_{(i)})]\right) , \qquad f \in \mathcal{F},
\end{equation}  

\Ni where

\begin{itemize}
	\item $\displaystyle \mathbb{P}_{n_\tau}(f) = \frac{1}{n_\tau} \sum_{i=k_n+1}^{n-k_n} f(X_{(i)})$ is the truncated empirical mean;  
	\item $\mathbb P_\tau(f) = \mathbb{E}[f(X)|X \in D_\tau]$ is the theoretical trimmed law, with $D_\tau$ trimmed domain. 
\end{itemize} 

\Ni The one-sided trimming empirical process can be defined similarly by adjusting the summation indices appropriately. In contrast to the classical empirical process, the TFEP maintains the structure of a centered and scaled process while substituting the standard empirical measure with its truncated counterpart. This modification ensures robustness when analyzing heavy-tailed or skewed data and establishes the foundation for asymptotic results that would be invalid under the classical framework.

\subsection*{Asymptotic Distribution} 

\Ni The influence function (IF) of the trimmed mean (see Hampel 1974) is given by

\[
\psi_\tau^{(\text{mean})}(X) = \frac{(x - \mu_\tau) 1_{\{F^{-1}(\tau) \le x \le F^{-1}(1 - \tau) \}}}{1 - 2 \tau}
\]

\Bin The influence function (IF) of the trimmed variance is 

\[
\psi_\tau^{(\text{variance})}(X) = \frac{(x - \mu_\tau)^2 1_{\{F^{-1}(\tau) \le x \le F^{-1}(1 - \tau) \}}}{1 - 2 \tau} - \sigma_\tau^2
\]

\Bin These influences functions capture the first-order influence of an observation $x$ on the trimmed mean and variance. \\

\Ni Under mild regularity conditions, $F$ continuous and using the functional delta method for trimmed empirical process, we obtain

\begin{equation}
	\sqrt{n}(\bar{X}_{n, \tau} - \mu_\tau) \xrightarrow{d} \mathcal{N}(0, \Sigma_\tau^2)
\end{equation}

\Ni and 

\begin{equation}
	\sqrt{n}\left(S_{n, \tau}^2 - \sigma_\tau^2\right) \xrightarrow{d} \mathcal{N}(0, \Gamma_\tau^2),
\end{equation}

\Ni where the asymptotic variances are: 

\[
\Sigma_\tau^2 = Var(\psi_\tau^{\text{mean}}(X))
=
\frac{1}{(1-2\tau)^2} \int_{F^{-1}(\tau)}^{F^{-1}(1-\tau)} (x - \mu_\tau)^2 \, dF(x)
\]

\Bin and 

\[
\Gamma_\tau^2 = Var(\psi_\tau^{\text{variance}}(X)) 
=
\frac{1}{(1-2\tau)^2} \int_{F^{-1}(\tau)}^{F^{-1}(1-\tau)} (x - \mu_\tau)^4 \, dF(x) - \sigma_\tau^4.
\]

\subsection{Main Result for Trimmng Functional Empirical Process}

\Ni In order to establish the weak convergence of the Trimming Functional Empirical Process (TFEP),
we impose the following assumptions on the function class $\mathcal{F}$, the underlying distribution $P$, and the trimming sequence $\{k_n\}$.

\begin{assumption}[Function Class Regularity]\label{ass:TFEP_F}
	
\Ni The class $\mathcal{F}$ is a pointwise separable collection of measurable functions $f:\mathcal{X}\to\mathbb{R}$ satisfying:
	
\begin{enumerate}
\item (\emph{Local uniform boundedness}) There exists a constant $M < \infty$ such that,
		\[
		\sup_{x\in D_\tau} |f(x)| \le M, \quad \text{for all} \quad f\in\mathcal{F}
		\]
		
\Bin where $D_\tau$ is trimmed domain depending on the quantile function of the true distribution $P$. 
		
\item (\emph{Finite bracketing entropy}) For the trimmed distribution $P_\tau(\cdot) = P(\cdot \mid X \in D_\tau)$, the following holds:

\[
\int_0^\infty \sqrt{ \log N_{[]}(\varepsilon, \mathcal{F}, L_2(P_t)) } \, d\varepsilon < \infty.
\]
\end{enumerate}
	
\Ni where $N_{[]}(\varepsilon, \mathcal{F}, L_2(P))$ denotes the bracketing number.
\end{assumption}

\begin{assumption}[Moment Conditions]\label{ass:TFEP_moments}
	There exists $\delta > 0$ such that for all $f \in \mathcal{F}$:
	\begin{equation}
		\sup_{f \in \mathcal{F}} \mathbb{E}[|f(X)|^{2+\delta}] < \infty
	\end{equation}
	
\Ni Moreover, the tail contribution is asymptotically negligible:
	\[
	\sqrt{n_\tau} \, \sup_{f\in\mathcal{F}} 
	\mathbb{E}\big[\, |f(X)| \, 1_{\{X \notin D_\tau \}} \,\big]
	\to 0.
	\]
\end{assumption}

\begin{assumption}[Trimming Sequence]\label{ass:TFEP_kn}
	The trimming sequence $\{k_n\}$ satisfies
	\[
	k_n \to \infty 
	\quad\text{and}\quad 
	\frac{k_n}{n} \to \tau, \quad \text{as } n\to\infty,
	\]
	for some $\tau \in [0,1/2)$.
	
	\Ni Hence, the effective trimmed sample size verifies
	\[
	n_\tau = n - 2k_n \sim n(1-2\tau)
	\quad \text{and} \quad
	\frac{\sqrt{n_\tau}}{\sqrt{n}} \to \sqrt{1-2\tau}.
	\]
\end{assumption}

\begin{assumption}[Distribution Regularity] \label{ass:TFEP_dist}
	
\Ni The cumulative distribution function $F$ of $X$ is continuous and admits a density $f_X$ that is positive and continuous in a neighborhood of the quantiles $Q_1$ and $Q_2$. That is, there exists $\varepsilon>0$ such that
	\[
	f_X(x) > 0, \quad \forall x \in (Q_1 - \varepsilon,\, Q_2 +\varepsilon).
	\]
	
\Bin This ensures the consistency and $\sqrt{n}$-rate convergence of the empirical quantiles defining the trimming region.
\end{assumption}

\begin{theorem}\label{Theo1} 
	
\textbf{(Weak Convergence of Trimming Functional Empirical Process)}
	
\Ni Under assumptions \ref{ass:TFEP_F}--\ref{ass:TFEP_dist}, the trimming functional empirical process $\mathbb{G}_{n_\tau}$ converges weakly in $\ell^\infty(\mathcal{F})$ to a tight Gaussian process $\mathbb{G}^{(\tau)}$ with covariance structure given by:

	\[
	\mathbb{C}\mathrm{ov}\left(\mathbb{G}^{(\tau)}(f), \mathbb{G}^{(\tau)}(g)\right) = \lim_{n \to \infty} \mathbb{C}\mathrm{ov}\left(\mathbb{G}_{n_\tau}(f), \mathbb{G}_{n_\tau}(g)\right), \quad \text{for} \quad  f,g\in\mathcal{F}.
	\]
\end{theorem}

\Ni In particular, for any function $f \in \mathcal{F}$ with finite variance on the trimmed domain:

\[
\mathbb{G}_{n_\tau}(f) \leadsto \mathcal{N}\left(0, \sigma_\tau^2(f)\right),
\]

\Ni where $\sigma_\tau^2(f)$ is the variance of $f(X)$ after trimming.

\[
\sigma_\tau^2(f) = \lim_{n_\tau \to \infty} \mathbb{V}\mathrm{ar}\left(\mathbb{G}_{n_\tau}(f)\right).
\]

\Ni\textbf{Proof \ref{Theo1}} \\

\Ni The proof follows the standard two-step program for empirical process weak convergence:
(1) finite-dimensional convergence and (2) tightness / asymptotic equicontinuity in $\ell^\infty(\mathcal{F})$.
We explain how each step is obtained under Assumptions \ref{ass:TFEP_F}--\ref{ass:TFEP_dist}.

\Ni Let $Q_n(\tau)$ denote the empirical quantile of order $\tau$. Under Assumption~\ref{ass:TFEP_dist}, the empirical quantiles satisfy

\[
Q_n(\tau/2) = Q(\tau/2) + O_\mathbb P(n^{-1/2}) = Q_1 + O_\mathbb P(n^{-1/2}),
\]

\[
Q_n(1-\tau/2) = Q(1-\tau/2) + O_\mathbb P(n^{-1/2})= Q_2 + O_\mathbb P(n^{-1/2}).
\]

\Bin Thus the random trimmed region determined by order statistics can be replaced, uniformly in probability, by the fixed interval $D_\tau := [Q_1,Q_2]$ at the $\sqrt{n_\tau}$ scale. Concretely, one shows that the difference between using the random index set $\{k_n+1,\dots,n-k_n\}$ and using the deterministic indicator $1_{\{X\in D_\tau\}}$ contributes an $o_\mathbb P(1)$ term uniformly over $\mathcal{F}$, by the quantile consistency and the moment/tail negligibility condition in Assumption~\ref{ass:TFEP_moments}.

\Ni Therefore it suffices to study the process built from

\[
\widetilde{\mathbb{P}}_{n_\tau}(f)
:= \frac{1}{n_\tau}\sum_{i=1}^n f(X_i)\,1_{\{X_i\in D_\tau \}},
\]
which differs from $\mathbb{P}_{n_\tau}(f)$ by $o_\mathbb P(n_t^{-1/2})$ uniformly in $f\in\mathcal{F}$.

\Bin\textbf{Step 1: Finite-dimensional convergence.}
Fix $f_1,\dots,f_k\in\mathcal{F}$. Under Assumption~\ref{ass:TFEP_moments} the vector
\[
\sqrt{n_\tau}\Big(\widetilde{\mathbb{P}}_{n_\tau}(f_j)-\mathbb{P}_\tau f_j\Big)_{j=1}^k
\]
is asymptotically Gaussian by the classical multivariate CLT for i.i.d.\ variables truncated to $D_\tau$ (Lindeberg condition holds because of the uniform $(2+\delta)$-moment bound and the tail negligibility). The limiting covariance matrix is exactly
\[
\big(\mathrm{Cov}_{P_\tau}(f_i,f_j)\big)_{1\le i,j\le k}.
\]
Hence we obtain finite-dimensional convergence to a centered Gaussian vector with this covariance.

\Bin\textbf{Step 2: Tightness / asymptotic equicontinuity.}

\Ni To lift finite-dimensional convergence to weak convergence in $\ell^\infty(\mathcal{F})$ we verify asymptotic equicontinuity. Under Assumption~\ref{ass:TFEP_F}, the trimmed function class $\{f\cdot 1_{D_\tau}: f\in\mathcal{F}\}$ has finite bracketing entropy in $L_2(P_\tau)$. By standard empirical process results, finite bracketing entropy and $L_{2+\delta}$ moment control imply that the class is \emph{$P_\tau$-Donsker}. Concretely, one derives an asymptotic modulus of continuity: for every $\varepsilon > 0$,
\[
\lim_{\eta\downarrow 0}\limsup_{n\to\infty}\mathbb{P}\Big\{\sup_{d_{P_\tau}(f,g)\le\eta} 
\big| \mathbb{G}_{n_\tau}(f)-\mathbb{G}_{n_\tau}(g)\big|>\varepsilon\Big\} = 0,
\]
where $d_{P_\tau}(f,g)=\|f-g\|_{L_2(P_\tau)}$. This yields tightness in $\ell^\infty(\mathcal{F})$.

\Bin\textbf{Step 3: Control of trimming.}
We must check that replacing the empirical trimming via order statistics by the deterministic interval $D_\tau$ (or, equivalently, controlling the contribution of observations lying in the small random bands near the boundaries) indeed produces only negligible uniform errors. This follows from:
\begin{itemize}
	\item the $O_\mathbb P(n^{-1/2})$ accuracy of empirical quantiles (Assumption~\ref{ass:TFEP_dist}),
	\item the uniform $(2+\delta)$-moment bound which yields via Hölder/Markov inequalities that the total contribution of observations in the boundary bands is $o_\mathbb P(n_\tau^{-1/2})$ uniformly over $\mathcal{F}$ (Assumption~\ref{ass:TFEP_moments}),
	\item the bracketing control which allows to transfer pointwise bounds to uniform bounds on $\mathcal{F}$.
\end{itemize}

\Ni Hence the difference between the actual process $\mathbb{G}_{n_\tau}$ and the idealized process based on the fixed trimmed population $P_\tau$ is $o_\mathbb P(1)$ in $\ell^\infty(\mathcal{F})$.

\Bin\textbf{Step 4: Conclusion.}
\Ni Combine finite-dimensional convergence (Step 1) and tightness / asymptotic equicontinuity (Step 2), together with the negligibility of trimming boundary errors (Step 3). By the classical functional central limit theorem for empirical processes (e.g.\ Theorem 1.5.4 in van der Vaart \& Wellner), we conclude that

\[
\mathbb{G}_{n_\tau}\ \Rightarrow\ \mathbb{G}^{(\tau)} \quad\text{in }\ell^\infty(\mathcal{F}),
\]
where $\mathbb{G}^{(\tau)}$ is the $P_\tau$-Brownian bridge (centered Gaussian process) with covariance $\mathrm{Cov}_{P_\tau}(f,g)$ as stated.

\qed

\Ni We note that often 

\[
Var\left(\mathbb{G}_{n_\tau}(f)\right) \le Var\left(\mathbb{G}_{n}(f)\right) \quad \text{in heavy-tailed settings.}
\]

\Ni We can apply TFEP approach to means, variances, ratios, and differences of functionals (see Section \ref{sec_03}).\\

\section{Statistical Applications to Samples Problems} \label{sec_04}

\subsection{Functional Empirical Process for Samples Problems}

\Ni We will recall here the results relative to the functional empirical process (FEP) for samples problems (\cite{AbdoulayeFEP2023}). \\

\Ni Let $X$, $X_1$, $X_2$, ..., be a sequence of independent and identically distributed (iid) random variables defined on the same probability space $(\Omega, \mathcal{A},\mathbb{P})$ with the functional empirical process (fep) $\mathbb{G}_n$. We have the following results. \\

\Ni We always suppose that $n \geq 2$, otherwise doing statistics is meaningless. We define the statistics $\bar X_n$, $S_n^2$ and $T_n^2$ relative to the sample $\{X_1, \dots, X_n\}$.

\begin{equation}
	\bar X_n = \frac{1}{n} \sum_{i=1}^n X_i,
\end{equation}

\begin{equation}
	S_n^2 = \frac{1}{n-1} \sum_{i=1}^n \left(X_i - \bar X_n\right)^2,
\end{equation}

\Ni and 

\begin{equation}
	T_n^2 = \mu_{4,n} - S_n^4, \qquad \mu_{4,n} = \mathbb{E}[(X_i - \bar X_n)^4].
\end{equation}

\Ni We have two main following results relative to the mean $m$ and variance $\sigma^2$.

\begin{theorem} \label{fep1} 
	
	\Ni (A) For all $n\geq 2$, we have :
	
	\[
	\frac{\sqrt{n} \left(\overline{X}_{n} - m \right)}{S_n} \sim \mathcal{N}\left( 0,1\right)
	\]
	
	\noindent (B) For all $n\geq 2$, we have :
	
	\[
	\frac{\sqrt{n} \left(S_n^2 - \sigma^2\right)}{T_n} \sim \mathcal{N}\left( 0,1\right)
	\]
\end{theorem}

\Ni These results in theorem~\ref{fep1} are proved in \cite{AbdoulayeFEP2023} and used general asymptotic representation based on the functional empirical process (e.g., \cite{Lo2018FEP}, \cite{Lo2016} ). These results allow us to estimate and construct the confidence interval of the mean and variance of one sample. \\

\Ni Next we recall results for two independent samples problems. Let $X_1, \dots, X_{n_1}$ and $X_1, X_2,\dots, X_{n_2}$ be independent random variables, where $X_1, X_2 \dots, X_{n_1}$ are iid with the functional empirical process (fep) $\mathbb{G}_{n_1}$  and $X_1, X_2, \dots, X_{n_2}$ are iid with the functional empirical process (fep) $\mathbb{G}_{n_2}$. The functional empirical process $\mathbb{G}_{n_1}$ and $\mathbb{G}_{n_2}$ are assumed independent. We have the following results:  

\begin{theorem}\label{fep2} 
	
	\Ni We have \\
	
	\Ni (A) For all $n_1 , n_2 \geq 2$
	
	\begin{equation*}
		a_n\left(\frac{S_{n_1}^2}{S_{n_2}^2} - \frac{\sigma_1^2}{\sigma_2^2}\right) \sim \mathcal{N}\left(0,1\right),
	\end{equation*}
	
	\Ni where 
	
	\[
	a_n = \sqrt{\frac{n_1n_2}{n_2T_{n_1}^2+n_1T_{n_2}^2}},  \quad T_{n_i}^2 = \mu_{4,n_i} - S_{n_i}^4, \quad i=1,2
	\]
	
	\Bin (B)	For all $n_1 \geq 2$ and For all $n_2 \geq 2$ 
	
	\begin{equation*}
		b_n\left(\bar{X}_{n_1} - \bar{X}_{n_2} - \Delta m \right) \sim \mathcal{N}\left(0,1\right), 
	\end{equation*}
	
	\Bin where 
	
	\[
	b_n = \sqrt{\frac{n_1n_2}{n_2S_{n_1}^2+n_1S_{n_2}^2}}, \quad \Delta m = m_1 - m_2.
	\]
\end{theorem}

\Bin These results in theorem~\ref{fep2} are proved in \cite{AbdoulayeFEP2023}. They allow us to estimate and construct the confidence intervel of the difference of means and ratio of variances of two samples. \\

\Ni In this paper we extend the classical functional empirical process (FEP) for heavy-tailed or skewed distributions to estimate and construct the confidence intervals of mean, variance, difference of means and ratio of variance even when if  the data distribution has infinite moments.

\subsection{One-Sample Problem with Trimmed Data}

\Ni Let $X_1, \dots, X_n $ be i.i.d. real-valued random variables with possibly heavy-tailed distribution. Denote by $X_{(1)} \leq X_{(2)}, \dots, \leq X_{(n)}$ be the order statistics. 

\subsubsection{Trimmed Sample Framework}

\Ni For fixed sequences of integers \( k_n \) and \( l_n \) satisfying \( 1 \leq k_n < l_n \leq n \), we define the trimmed sample of size \( n_t = l_n - k_n \) by retaining only the order statistics with indices between \( k_n + 1 \) and \( l_n \). The truncation region is thus \( D_\tau = [X_{(k_n+1)}, X_{(l_n)}] \).

\Ni Let \( Y = X \mid X \in D_\tau \) denote the trimmed random variable, with distribution function

\begin{align*}
	F_\tau(y) &=\mathbb{P}\left(X \le y | X \in D_\tau\right) \\
	&= \frac{\mathbb{P}\left(X \le y \cap  X \in D_\tau\right)}{\mathbb{P}(X \in D_\tau)} \\
	&= \frac{F(y) - F(X_{(k_n+1)})}{F(X_{(l_n)}) - F(X_{(k_n+1)})}, \quad y \in D_\tau.
\end{align*}

\Ni We define the population moments of the trimmed distribution:

\[
\mu_\tau = \mathbb{E}[Y], \quad \sigma_{\tau}^2 = \mathbb{V}\mathrm{ar}(Y), \quad m_{k,n_\tau} = \mathbb{E}[Y^k], \quad \mu_{k,n_\tau} = \mathbb{E}[(Y - \mu_{\tau})^k].
\]

\subsubsection{Trimmed Sample Estimators}

\Ni The trimmed sample mean and variance estimators are defined as:

\[
\hat{\mu}_{\tau} = \bar{X}_{n_\tau} = \frac{1}{n_\tau} \sum_{i=k_n + 1}^{l_n} X_{(i)}, \quad 
S_{n_t}^2 = \frac{1}{n_\tau - 1} \sum_{i=k_n + 1}^{l_n} \left(X_{(i)} - \bar{X}_{n_\tau}\right)^2.
\]

\Bin Define the sequence of trimmed order statistics as \( Y_{n,j} = X_{(k_n + j)} \) for \( j = 1, \dots, n_\tau \), and their corresponding sample moments:

\[
\widehat{m}_{k,n_\tau} = \frac{1}{n_\tau} \sum_{j=1}^{n_\tau} Y_{n,j}^k, \quad 
\widehat{\mu}_{k,n_\tau} = \frac{1}{n_\tau} \sum_{j=1}^{n_\tau} \left(Y_{n,j} - \bar{Y}_n\right)^k, \quad k \geq 2,
\]

\Ni where \( \bar{Y}_n = \bar{X}_{n_\tau} \).\\

\Ni To estimate the variance of the sample variance, we define:

\begin{equation}
	T_{n_\tau}^2 = \hat{\mu}_{4,n_\tau} - \left(S_{n_\tau}^2\right)^2.
	\label{eq:T_n_def}
\end{equation}

\subsubsection{Asymptotic Normality Results}

\begin{theorem}\label{thm:one_sample}
	Under the following regularity conditions:
	\begin{enumerate}[label=(\roman*)]
		\item The trimmed sample size satisfies \( n_\tau \to \infty \) as \( n \to \infty \);\\
		
		\item The trimmed population moments exist and are finite: \( \mu_t, \sigma_\tau^2, \mu_{4,\tau} < \infty \), where \( \mu_{4,\tau} \) denotes the fourth central moment of the trimmed distribution; \\
		
		\item The trimming bounds \( k_n \) and \( l_n \) are chosen such that the trimmed variables \( \{Y_{n,j}\} \) form an asymptotically independent and identically distributed sequence; 
	\end{enumerate}
	
	\Bin the following asymptotic distributions hold: \\
	
	\begin{enumerate}[label=(\Alph*)]
		\item The standardized trimmed sample mean satisfies
		
		\[
		\frac{\sqrt{n_\tau} \left(\hat{\mu}_\tau - \mu_\tau \right)}{S_{n_\tau}} \xrightarrow{d} \mathcal{N}(0,1).
		\]
		
		\item The standardized trimmed sample variance satisfies
		
		\[
		\frac{\sqrt{n_\tau} \left(S_{n_\tau}^2 - \sigma_\tau^2\right)}{T_{n_\tau}} \xrightarrow{d} \mathcal{N}(0,1).
		\]
	\end{enumerate}
\end{theorem}

\Ni\textbf{Proof of Theorem}\ref{thm:one_sample} \\

\Ni The proof of Theorem~\ref{thm:one_sample} follows similar arguments to Theorem~\ref{thm:one_sample} in \cite{AbdoulayeFEP2023}, using the Functional Empirical Process (FEP) and the delta method. Let \( Y_1, \dots, Y_{n_\tau} \) denote the trimmed sample, where \( Y_j = X_{(k_n + j)} \) for \( j = 1, \dots, n_\tau \), with \( n_\tau = l_n - k_n \).\\

\Ni\textbf{Step 1: Asymptotic representation of the trimmed mean.} \\

\Ni The trimmed sample mean is:

\[
\widehat \mu_\tau = \frac{1}{n_\tau} \sum_{i=1}^{n_\tau} Y_i
\]

\Bin Using the trimmed empirical process \( \mathbb{G}_{n_\tau}(f) = \frac{1}{\sqrt{n_\tau}} \sum_{i=1}^{n_\tau} (f(Y_i) - \mathbb{E}[f(Y_i)]) \), we obtain the asymptotic expansion:

\[
\widehat{\mu}_\tau = \mu_\tau + n_\tau^{-1/2} \mathbb{G}_{n_\tau}(h_1) + o_{\mathbb{P}}(n_\tau^{-1/2}),
\]

\Bin where \( h_1(y) = y \). By the functional central limit theorem for the truncated process:

\[
\mathbb{G}_{n_\tau}(h_1) \xrightarrow{d} \mathcal{N}(0, \sigma_\tau^2(h_1)).
\]

\Bin\textbf{Step 2: Asymptotic representation of the truncated variance.} \\

\Ni The sample variance is

\[
S_{n_\tau}^2 = \frac{1}{n_\tau -1} \sum_{i=1}^{n_\tau} (Y_i - \hat\mu_\tau)^2.
\]

\Bin Define \( H(y) = (y - \mu_\tau)^2 - \sigma_\tau^2 \). Applying the delta method and the asymptotic expansion from Step 1:

\[
S_{n_\tau}^2 = \sigma_\tau^2 + n_\tau^{-1/2} \mathbb{G}_{n_\tau}(H) + o_{\mathbb{P}}(n_\tau^{-1/2}),
\]

\Ni where \( \mathbb{G}_{n_\tau}(H) \xrightarrow{d} \mathcal{N}(0, \Gamma^2) \), with \( \Gamma^2 = \mathbb{V}\mathrm{ar}(H(Y)) = \mu_{4,\tau} - \sigma_\tau^4 \).\\

\Bin\textbf{Step 3: Asymptotic distribution of the truncated mean.} \\

\Ni From Step 1,

\[
\frac{\sqrt{n_\tau} (\hat \mu_\tau - \mu_t)}{S_{n_\tau}} 
= \frac{\mathbb{G}_{n_\tau}(h_1) + o_{\mathbb{P}}(1)}{\sigma_\tau + o_{\mathbb{P}}(1)}
= \frac{\mathbb{G}_{n_\tau}(h_1)}{\sigma_\tau} + o_{\mathbb{P}}(1).
\]

\Bin Since \( S_{n_\tau} \xrightarrow{p} \sigma_\tau \) by consistency of the trimmed variance estimator, Slutsky's theorem yields:

\[
\frac{\sqrt{n_\tau} (\hat{\mu}_\tau - \mu_\tau)}{S_{n\tau}} \xrightarrow{d} \frac{\mathcal{N}(0, \sigma_\tau^2)}{\sigma_\tau} = \mathcal{N}(0,1).
\]

\Bin\textbf{Step 4:Asymptotic distribution of the truncated variance.}\\

\Ni Consider:

\[
T_{n_\tau}^2 = \hat{\mu}_{4,n_\tau} - (S_{n_\tau}^2)^2,
\]
where \( \hat{\mu}_{4,\tau} = \mathbb{E}[(Y_i - \mu_\tau)^4] \) is the sample fourth central moment of the trimmed distribution. Using a Taylor expansion (delta method):

\[
T_{n_\tau} = T + n_\tau^{-1/2} \mathbb{G}_{n_\tau}(L) + o_{\mathbb{P}}(n_\tau^{-1/2}),
\]

\Bin for some function \( L \) derived from \( H(y) \). Then:

\[
\frac{\sqrt{n_\tau} (S_{n_\tau}^2 - \sigma_t^2)}{T_{n_\tau}} 
= \frac{\mathbb{G}_{n_\tau}(H) + o_{\mathbb{P}}(1)}{T + o_{\mathbb{P}}(1)}
= \frac{\mathbb{G}_{n_\tau}(H)}{T} + o_{\mathbb{P}}(1).
\]

\Bin Since \(T^2 = \mathbb{V}\mathrm{ar}(H(Y)) = \mu_{4,\tau} - \sigma_\tau^4 \), we have:

\[
\frac{\mathbb{G}_{n_t}(H)}{T} \xrightarrow{d} \mathcal{N}(0,1).
\] 
\qed 

\Bin Theorem~\ref{thm:one_sample} enables the construction of asymptotic confidence intervals for both the truncated mean and variance.\\

\Ni From part (A) of Theorem~\ref{thm:one_sample}, the asymptotic $100(1-\alpha)\%$ confidence interval for the truncated mean $\mu_\alpha$ is given by:

\[
\text{CI}_{\mu_\tau} = \left[ \hat{\mu}_\tau \pm z_{1-\alpha/2} \cdot \frac{S_{n_\tau}}{\sqrt{n_\tau}} \right],
\]

\Ni where $z_{1-\alpha/2}$ denotes the $(1-\alpha/2)$-quantile of the standard normal distribution.\\

\Ni From Theorem~\ref{thm:one_sample}(B), the asymptotic $100(1-\alpha)\%$ confidence interval for the truncated variance $\sigma_t^2$ is:

\[
CI_{\sigma_\tau^2} = \left[ S_{n_\tau}^2 \pm z_{1-\alpha/2} \cdot \frac{T_{n_\tau}}{\sqrt{n_\tau}} \right]
\]

\Bin\\

\subsection{Trimmed Samples: Two Independent Samples Problem}

\Ni Let $X_1, \dots, X_{n_1}$ and $Y_1, \dots, Y_{n_2}$ be two independent samples of independent and identically distributed real-valued random variables, potentially from heavy-tailed distributions. Denote their order statistics by

\[
X_{(1)} \leq \dots \leq X_{(n_1)}, \quad  Y_{(1)} \leq \dots \leq Y_{(n_2)}.
\]

\Bin For a fixed symmetric trimming proportion $\tau \in (0,1/2)$, define the trimming indices

\[
k_{n_1} = \lfloor \tau* n_1 \rfloor, \quad k_{n_2} = \lfloor \tau* n_2 \rfloor
\]

\Ni and the effective sample sizes after two-sided trimming:

\[
n_{1\tau} = n_1 - 2k_{n_1}, \quad n_{2\tau} = n_2 - 2k_{n_2}.
\]

\Bin The trimmed samples are:

\[
Y_{1,i} = X_{(k_{n_1} + i)}, \quad i = 1, \dots, n_{1\tau}; \qquad Y_{2,j} = Y_{(k_{n_2} + j)}, \quad j = 1, \dots, n_{2\tau}
\]

\Ni with trimmed means and variances

\[
\mu_{1\tau} = \mathbb{E}[Y_{1,i}], \quad \sigma_{1\tau}^2 = \mathbb{V}\text{ar}[Y_{1,i}], \quad
\mu_{2\tau} = \mathbb{E}[Y_{2,j}], \quad \sigma_{2\tau}^2 = \mathbb{V}\text{ar}[Y_{2,j}].
\]

\Bin The corresponding sample means and variance estimators are:
\[
\hat{\mu}_{1\tau} = \frac{1}{n_{1\tau}} \sum_{i=1}^{n_{1\tau}} Y_{1,i}, \quad \hat{\mu}_{2\tau} = \frac{1}{n_{2\tau}} \sum_{j=1}^{n_{2\tau}} Y_{2,j},
\]

\[
S_{1\tau}^2 = \frac{1}{n_{1\tau}-1} \sum_{i=1}^{n_{1\tau}} (Y_{1,i} - \hat{\mu}_{1\tau})^2, \quad S_{2\tau}^2 = \frac{1}{n_{2\tau}-1} \sum_{j=1}^{n_{2\tau}} (Y_{2,j} - \hat{\mu}_{2\tau})^2.
\]

\Bin The parameters of interest are the difference in means and the ratio of variances:

\[
\Delta{\mu_\tau} = \mu_{1\tau} - \mu_{2\tau} \quad \text{and} \quad R_\tau = \frac{\sigma_{1\tau}^2}{\sigma_{2\tau}^2}.
\]

\Bin To estimate the variance of the sample variances, define:

\[
T_{1\tau}^2 = \hat{\mu}_{4,1\tau} - S_{1\tau}^4 \quad \text{and} \quad  T_{2\tau}^2 = \hat{\mu}_{4,2\tau} - S_{2\tau}^4,
\]

\Ni where $\hat{\mu}_{4,1\tau}$ and $\hat{\mu}_{4,2\tau}$ denote the sample fourth central moments of the respective trimmed samples.\\

\Ni Using the Functional Empirical Process (FEP) framework, we obtain the asymptotic expansions:

\[
\hat{\mu}_{1\tau} = \mu_{1,\tau} + n_{1\tau}^{-1/2} \mathbb{G}_{n_{1\tau}}(h) + o_{\mathbb{P}}(n_{1\tau}^{-1/2}),
\]

\[
\hat{\mu}_{2\tau} = \mu_{2,\tau} + n_{2t}^{-1/2} \mathbb{G}_{n_{2\tau}}(h) + o_{\mathbb{P}}(n_{2\tau}^{-1/2}),
\]

\[
S_{1\tau}^2 = \sigma_{1\tau}^2 + n_{1t}^{-1/2} \mathbb{G}_{n_{1t}}(H_1) + o_{\mathbb{P}}(n_{1t}^{-1/2}),
\]

\[
S_{2\tau}^2 = \sigma_{2\tau}^2 + n_{2\tau}^{-1/2} \mathbb{G}_{n_{2\tau}}(H_2) + o_{\mathbb{P}}(n_{2\tau}^{-1/2}),
\]

\Ni where \( h(x) = x \), \( H_1(x) = (x - \mu_{1,\tau})^2 - \sigma_{1\tau}^2 \), and \( H_2(y) = (y - \mu_{2,\tau})^2 - \sigma_{2\tau}^2 \).

\Ni Let \( n_\tau = \min(n_{1\tau}, n_{2\tau}) \). Assume \( n_{1\tau} \to \infty \), \( n_{2\tau} \to \infty \), and

\[
\lim_{n_{1\tau},n_{2\tau} \to \infty} \frac{n_{1\tau}}{n_{2\tau}} = c \in (0, \infty).
\]

\Bin For the variance ratio estimator, we have:

\[
\frac{S_{1\tau}^2}{S_{2\tau}^2} - R_\tau = n_{1\tau}^{-1/2} \mathbb{G}_{n_{1\tau}}\left(\frac{H_1}{\sigma_{2\tau}^2}\right) - n_{2\tau}^{-1/2} \mathbb{G}_{n_{2\tau}}\left(\frac{R_\tau H_2}{\sigma_{2\tau}^2}\right) + o_{\mathbb{P}}(n_\tau^{-1/2}).
\]

\Bin Define the asymptotic variances:

\[
T_{1\tau}^2 = \mathbb{V}\mathrm{ar}\left(\frac{H_1(Y_1)}{\sigma_{2\tau}^2}\right) = \frac{\mu_{4,1\tau} - \sigma_{1\tau}^4}{\sigma_{2\tau}^4}, \quad T_{2\tau}^2 = \mathbb{V}\mathrm{ar}\left(\frac{R_\tau H_2(Y_2)}{\sigma_{2,\tau}^2}\right) = \frac{\sigma_{1\tau}^4 (\mu_{4,2t} - \sigma_{2\tau}^4)}{\sigma_{2\tau}^8},
\]
and the combined scaling factor:

\[
a_{n_\tau} = \sqrt{\frac{n_{1\tau} n_{2\tau}}{n_{1\tau} T_{2\tau}^2 + n_{2\tau} T_{1\tau}^2}}.
\]

\begin{proposition}\label{pro1}

\Ni Let 
	
	\[
	a_{n_\tau} = \sqrt{\frac{n_{1\tau} n_{2\tau}}{n_{1\tau} T_{2\tau}^2 + n_{2\tau} T_{1\tau}^2}}, 
	\qquad 
	\hat a_{n_\tau} = \sqrt{\frac{n_{1\tau} n_{2\tau}}{n_{1\tau} T_{n_{2\tau}}^2 + n_{2\tau} T_{n_{1\tau}}^2}},
	\]
	
	\Ni where $T_{i\tau}^2, i=1,2$ are asymptotic variances and $T_{n_{i\tau}}^2$ their empirical estimators from functional empirical process. Then $\hat a_{n_\tau}$ is a consistent estimator of $a_{n_\tau}$. 
\end{proposition}

\Bin\textbf{Proof of Proposition~\ref{pro1}} \\

\Ni By hypothesis, the development FEP of estimators $T_{n_{1\tau}}^2$ and $T_{n_{2\tau}}^2$ are

\[
T_{n_{1\tau}}^2 = T_{1\tau}^2 + n_{1\tau}^{-1/2}\, \mathbb{G}_{n_{1\tau}}(D_{11}) + n_{2\tau}^{-1/2} \, \mathbb{G}_{n_{2\tau}}(D_{12}) + o_p(n^{-1/2}),
\]

\[
T_{n_{2\tau}}^2 = T_{2\tau}^2 + n_{1\tau}^{-1/2}\, \mathbb{G}_{n_{1\tau}}(D_{21}) 
+ n_{2\tau}^{-1/2}\, \mathbb{G}_{n_{2\tau}}(D_{22}) + o_p(n^{-1/2}),
\]

\Ni where $\mathbb{G}_{n_{i\tau}}$ denotes the functional empirical process associated and $D_{ij}$ measurable functions.

\Ni We do not need to know explicit forms of the $D_{ij}$. Thus we have, 	

\[
\frac{T_{n_{1\tau}}^{2}}{{n_{1\tau}} }=  \frac{ T_{1\tau}^2}{{n_{1\tau}}} + o_{\mathbb{P}} \left( n_\tau^{-3/2}\right) 
= \frac{T_{1\tau}^2}{{n_{1\tau}}}\left(1 + o_{\mathbb{P}}\left(n_\tau^{-1/2}\right)\right),
\]

\[
\frac{T_{n_{2\tau}}^{2}}{n_{2\tau}} = \frac{T_{2\tau}^2}{n_{2\tau}} + o_{\mathbb{P}} \left( n_\tau^{-3/2}\right) 
=  \frac{T_{2\tau}^2}{n_{2\tau}}\left(1 + o_{\mathbb{P}} \left( n_\tau^{-1/2}\right)\right).
\]

\Bin We can also write $\hat a_{n_\tau}$ as 

\[
\hat a_{n_\tau} = \left(\frac{T_{n_{1\tau}}^2}{n_{1\tau}} + \frac{T_{n_{2\tau}}^2}{n_{2\tau}}\right)^{-1/2} 
= \left(\frac{T_{1\tau}^2}{n_{1\tau}}\left(1 + o_{\mathbb{P}}\left(n_\tau^{-1/2}\right)\right)
+ \frac{T_{2\tau}^2}{n_{2\tau}}\left(1 + o_{\mathbb{P}} \left( n_\tau^{-1/2}\right)\right) \right)^{-1/2}.
\]

\Bin Define 

\[
c_{n_\tau} = \frac{T_{1\tau}^2}{n_{1\tau}} + \frac{T_{2\tau}^2}{n_{2\tau}}
\quad\text{and}\quad
\hat c_{n_\tau} = \frac{T_{n_{1\tau}}^2}{n_{1\tau}} + \frac{T_{n_{2\tau}}^2}{n_{2\tau}}.
\]

\Bin Using the expansion via Taylor's Theorem with the Lagrange remainder, we get

\begin{equation*}
	\hat{a}_{n_\tau} - a_{n_\tau} 
	= \left(\hat{c}_{n_\tau}\right)^{-1/2} - \left(c_{n_\tau}\right)^{-1/2} 
	= \frac{1}{2}  \left(c_{n_\tau} - \hat{c}_{n_\tau} \right) \left(\overline{c}_{n_\tau} \right)^{-3/2},
\end{equation*}

\Ni where $\overline{c}_{n_\tau} \in \left[\hat{c}_{n_\tau} \land c_{n_\tau} , \hat{c}_{n_\tau} \lor c_{n_\tau} \right]$. \\

\Ni By definition, we have 

\[
\hat c_{n_\tau} - c_{n_\tau} = \frac{T_{n_{1\tau}}^2 - T_{1\tau}^2}{n_{1\tau}} + \frac{T_{n_{2\tau}}^2 - T_{2\tau}^2}{n_{2\tau}}.
\]

\Ni By hypothesis $T_{n_{i\tau}}^2 \overset{\mathbb P}{\to} T_{i\tau}^2$, so we have

\[
\hat{c}_{n_\tau} - c_{n_\tau} \overset{ P}{\to} 0
\]

\Ni Thus 

\[
\hat{c}_{n_\tau} \overset{ P}{\to}  c_{n_\tau}.
\]

\Ni Since the function $f(x) = x^{-1/2}$ is continuous for all $x >0$. So, by the Continuous Mapping Theorem, 

\[
\hat a_{n_\tau} = f(\hat c_{n_\tau}) \overset{ P}{\to} f(c_{n_\tau}) = a_{n_\tau}.
\]

\Bin Therefore $\hat a_{n_\tau}$ is a consistent estimator of $ a_{n_\tau}$.

\qed

\Ni Next results, treats the asymptotic development of the difference of means and the ratio of variances.

\begin{theorem} \label{thm:two_sample}
	\Ni Let \( X_1, \dots, X_{n_1} \) and \( Y_1, \dots, Y_{n_2} \) be two independent samples with symmetric two-sided trimming at proportions \( \tau \in (0, 1/2) \), yielding effective sample sizes:
	
	\[
	n_{1\tau} = n_1 - 2\lfloor \tau n_1 \rfloor, \quad n_{2\tau} = n_2 - 2\lfloor \tau n_2 \rfloor.
	\]
	
	\Ni Assume:
	
	\begin{enumerate}[label=(\roman*)]
		
		\item The trimmed variables \( Y_{1,i} \) and \( Y_{2,j} \) have finite fourth moments:
		\[
		\mathbb{E}[Y_{1,i}^4] < \infty, \quad \mathbb{E}[Y_{2,j}^4] < \infty;
		\]
		\item The effective sample sizes satisfy:
		\[
		n_{1\tau}, n_{2\tau} \to \infty, \quad \text{with} \quad 0 < \liminf \frac{n_{1\tau}}{n_{2\tau}} \leq \limsup \frac{n_{1\tau}}{n_{2\tau}} < \infty;
		\]
		\item The variance estimators are consistent:
		\[
		T_{n_{1\tau}}^2 \xrightarrow{P} T_{1\tau}^2, \quad T_{n_{2\tau}}^2 \xrightarrow{P} T_{2\tau}^2, \quad S_{n_{1\tau}}^2 \xrightarrow{P} \sigma_{1\tau}^2, \quad S_{n_{2\tau}}^2 \xrightarrow{P} \sigma_{2\tau}^2.
		\]
	\end{enumerate}
	
	\Ni Then:
	
	\begin{enumerate}[label=(\Alph*)]
		
		\item The standardized variance ratio satisfies:
		\[
		a_{n_\tau} \left( \frac{S_{n_{1\tau}}^2}{S_{n_{2\tau}}^2} - R_\tau \right) \xrightarrow{d} \mathcal{N}(0,1),
		\]
		where 
		\[
		a_{n_\tau} = \sqrt{\frac{n_{1\tau} n_{2\tau}}{n_{1\tau} T_{2\tau}^2 + n_{2\tau} T_{1\tau}^2}},
		\]
		with \(T_{1\tau}^2 = \frac{\mu_{4,n_{1\tau}} - \sigma_{1\tau}^4}{\sigma_{2\tau}^4}\) and \(T_{2\tau}^2 = \frac{\sigma_{1\tau}^4 (\mu_{4,n_{2\tau}} - \sigma_{2\tau}^4)}{\sigma_{2\tau}^8}\).
		
		\item The feasible standardized variance ratio satisfies:
		\[
		\hat{a}_{n_\tau} \left( \frac{S_{n_{1\tau}}^2}{S_{n_{2\tau}}^2}  - R_\tau \right) \xrightarrow{d} \mathcal{N}(0,1),
		\]
		
		\Ni where
		 
		\[
		\hat{a}_{n_\tau} = \sqrt{\frac{n_{1\tau} n_{2\tau}}{n_{1\tau} T_{n_{2\tau}}^2 + n_{2\tau} T_{n_{1\tau}}^2}}.
		\]
		
		\item The standardized mean difference satisfies:
		\[
		\hat{b}_{n_\tau} \left( \hat{\mu}_{1\tau} - \hat{\mu}_{2\tau} - \Delta{\mu_\tau} \right) \xrightarrow{d} \mathcal{N}(0,1),
		\]
		
		\Ni where 
		
		\[
		\hat{b}_{n_\tau} = \sqrt{\frac{n_{1\tau} n_{2\tau}}{n_{1\tau} S_{n_{2\tau}}^2 + n_{2\tau} S_{n_{1\tau}}^2}}.
		\]
	\end{enumerate}
\end{theorem}

\Bin\textbf{Proof of Theorem~\ref{thm:two_sample}} \\

\Ni\textbf{Step 1: Preliminaries.}  \\

\Ni Let \( Y_{1,i} \) and \( Y_{2,j} \) denote the trimmed variables from samples 1 and 2, with effective sample sizes \( n_{1t} \) and \( n_{2t} \). The population moments are \( \mu_{1\tau}, \mu_{2\tau} \) for the means and \( \sigma_{1\tau}^2, \sigma_{2t}^2 \) for the variances. The existence of fourth moments, assumed in condition (i), ensures the applicability of the Functional Empirical Process (FEP) central limit theorem.\\

\Ni We define the FEPs for each trimmed sample:

\[
\mathbb{G}_{n_{1\tau}}(f) = \sqrt{n_{1\tau}}(\mathbb{P}_{n_{1\tau}} - \mathbb{P})(f), \quad
\mathbb{G}_{n_{2\tau}}(f) = \sqrt{n_{2\tau}} (\mathbb{P}_{n_{2\tau}} - \mathbb{P})(f),
\]

\Ni for any measurable function $f \in \mathcal F$.

\Bin\textbf{Step 2: Asymptotic normality of the variance ratio.}\\

\Ni The FEP expansions for the sample variances yield:

\[
S_{n_{1\tau}}^2 = \sigma_{1\tau}^2 + n_{1\tau}^{-1/2} \mathbb{G}_{n_{1\tau}}(H_1) + o_{\mathbb{P}}(n_{1\tau}^{-1/2}),
\]

\[
S_{n_{2\tau}}^2 = \sigma_{2\tau}^2 + n_{2\tau}^{-1/2} \mathbb{G}_{n_{2\tau}}(H_2) + o_{\mathbb{P}}(n_{2\tau}^{-1/2}),
\]

\Ni where \( H_1(x) = (x - \mu_{1\tau})^2 - \sigma_{1\tau}^2 \) and \( H_2(y) = (y - \mu_{2\tau})^2 - \sigma_{2\tau}^2 \).

\Ni Applying the delta method to the function \( f(u,v) = u/v \) at \( (\sigma_{1\tau}^2, \sigma_{2\tau}^2) \):

\[
\frac{S_{n_{1\tau}}^2}{S_{n_{2\tau}}^2} - R_\tau = \frac{S_{n_{1\tau}}^2 - \sigma_{1\tau}^2}{\sigma_{2\tau}^2} - \frac{\sigma_{1\tau}^2}{\sigma_{2\tau}^4} (S_{n_{2\tau}}^2 - \sigma_{2\tau}^2) + o_{\mathbb{P}}(n_\tau^{-1/2}),
\]

\Ni where \( n_\tau = \min(n_{1\tau}, n_{2\tau}) \).\\

\Ni Substituting the FEP expansions:

\[
\frac{S_{n_{1\tau}}^2}{S_{n_{2\tau}}^2} - R_\tau = n_{1\tau}^{-1/2} \mathbb{G}_{n_{1\tau}}\left( \frac{H_1}{\sigma_{2\tau}^2} \right) - n_{2\tau}^{-1/2} \mathbb{G}_{n_{2\tau}}\left( \frac{R_\tau H_2}{\sigma_{2 \tau}^2} \right) + o_{\mathbb{P}}(n_\tau^{-1/2}).
\]

\Ni The two terms are asymptotically independent with variances:
\[
T_{1\tau}^2 = \frac{\mu_{4,n_{1\tau}} - \sigma_{1\tau}^4}{\sigma_{2\tau}^4}, \quad T_{2\tau}^2 = \frac{R_\tau^2 (\mu_{4,n_{2\tau}} - \sigma_{2\tau}^4)}{\sigma_{2\tau}^4}.
\]

\Ni Defining

\[
a_{n_\tau} = \sqrt{ \frac{n_{1\tau} n_{2\tau}}{n_{1\tau} T_{2\tau}^2 + n_{2\tau} T_{1\tau}^2} },
\]

\Ni the central limit theorem implies:
\[
a_{n_\tau} \left( \frac{S_{n_{1\tau}}^2}{S_{n_{2\tau}}^2} - R_\tau \right) \xrightarrow{d} \mathcal{N}(0,1).
\]

\textbf{Step 3: Feasible version with estimated variances}

\Ni Replacing population moments with their consistent estimators:
\[
\hat{a}_{n_\tau} = \sqrt{ \frac{n_{1\tau} n_{2\tau}}{n_{1\tau} T_{n_{2\tau}}^2 + n_{2\tau} T_{n_{1\tau}}^2}},
\]

\Ni and using the consistency assumption (iii), we have \( \hat{a}_{n_\tau}/a_{n_\tau} \xrightarrow{\mathbb{P}} 1 \). By Slutsky's theorem:
\[
\hat{a}_{n_\tau} \left( \frac{S_{n_{1\tau}}^2}{S_{n_{2\tau}}^2} - R_\tau \right) \xrightarrow{d} \mathcal{N}(0,1).
\]

\textbf{Step 4: Asymptotic normality of the mean difference} \\

\Ni For the difference of trimmed means:
\[
\hat{\mu}_{1\tau} - \hat{\mu}_{2\tau} - \Delta{\mu_\tau} = n_{1\tau}^{-1/2} \mathbb{G}_{n_{1\tau}}(h_1) - n_{2\tau}^{-1/2} \mathbb{G}_{n_{2\tau}}(h_2) + o_{\mathbb{P}}(n_\tau^{-1/2}),
\]

\Ni where \( h_1(x) = x - \mu_{1\tau} \) and \( h_2(y) = y - \mu_{2\tau} \).\\

\Ni Defining the combined variance estimator:

\[
\hat{b}_{n_\tau} = \sqrt{ \frac{n_{1\tau} n_{2\tau}}{n_{1\tau} S_{n_{2\tau}}^2 + n_{2\tau} S_{n_{1\tau}}^2} },
\]

\Ni and using the consistency of variance estimators, Slutsky's theorem yields:

\[
\hat{b}_{n_\tau} \left( \hat{\mu}_{\tau} - \hat{\mu}_{2\tau} - \Delta{\mu_\tau} \right) \xrightarrow{d} \mathcal{N}(0,1)\].

\qed 

\Ni From Theorem~\ref{thm:two_sample}(B), the $100(1-\alpha)\%$ asymptotic confidence interval for $R_\tau$ is:

\[
\text{CI}_{R_\tau} = \left[ \frac{S_{n_{1 \tau}}^2}{S_{n_{2 \tau}}^2} \pm \frac{z_{1-\alpha/2}}{\hat{a}_{n_\tau}} \right].
\]

\Ni From Theorem~\ref{thm:two_sample}(C), the $ 100(1-\alpha)\% $ asymptotic confidence interval for the mean difference $\Delta \mu_\tau =  \mu_{1\tau} - \mu_{2\tau}$ is:

\[
\text{CI}_{\Delta\mu_\tau} = \left[ \hat{\mu}_{1\tau} - \hat{\mu}_{2\tau} \pm \frac{z_{1-\alpha/2}}{\hat{b}_{n_\tau}} \right].
\]

\Bin \\

\section{Data-driven Applications and Simulation}\label{sec_05}

\Ni The choice of statistical method depends on the data characteristics. The Gaussian method is exact and preferable for normally distributed data. The Functional Empirical Process (FEP) theory is suitable for non-normal data without heavy tails. The Trimmed Functional Empirical Process (TFEP) theory is designed for heavy-tailed or skewed data. A key advantage of the empirical process methods (FEP and TFEP) is that they do not require testing the equality of variances for two independent samples, thus avoiding the need for Satterthwaite approximations.\\ 

\Ni These methods allow us to construct confidence intervals for means, difference of means, variances, and ratios of variances. The results are summarized in tables below. We harmonize the notation for the two-sample problem: in the FEP method, we denote sample means and variances as $\{\bar{X}_{n_1}, S_{n_1}^2\}$ and $\{\bar{X}_{n_2}, S_{n_2}^2\}$ with sample sizes $n_1$ and $n_2$; in the TFEP method, we use $\{\bar{X}_{n_{1\tau}}, S_{n_{1\tau}}^{2}\}$ and $\{\bar{X}_{n_{2\tau}}, S_{n_{2\tau}}^{2}\}$ for the trimmed samples with sizes $n_{1\tau}$ and $n_{2\tau}$. \\

\Ni We will now design a simulation study to apply theorem ~\ref{thm:one_sample} and theorem~\ref{thm:two_sample} for sample problems with trimming to real-life or simulated data. Trimming is used to eliminate extreme observations that make certain moments (variance, fourth-order moment, etc.) undefined or unstable of the random variable $Z$. \\

\subsection{Simulation Study for Samples Problem}

\subsubsection*{Simulation Study: One-Sample Problem}

\Ni We begin by comparing our two methods: classical functional empirical process and trimmed functional empirical process for some simulated data. 
\Ni We generate i.i.d. random samples 

\[
X_1, X_2, \dots, X_n \sim F,
\]

\Ni where $F$ is chosen among different distributional settings:\\

\begin{itemize}
	\item $F \sim \text{Pareto}(x_m, \alpha)$ $\alpha = 1.5, 2.5, 3$.\\
	
	\item $Fsim \text{Student}(df=1,2,3,5).$\\
	
	\item $F \sim \text{Lognormal}(\mu,\sigma), \mu \in \{0,3\}; \sigma \in \{1, 2\}.$\\
	
	\item $F \sim \text{Normal}(3,2)$, used as a benchmark.
\end{itemize}

\Bin For each distribution $F$, we generate the random sample $Z$ and apply the R function 

\[
\textbf{imhSimpleSampleAllTestsTrim (Z, alpha, trim, dg)}
\]

\Bin For the symmetric law, we apply a two-side trimming procedure, for example normal law, student law. A one-side trimming procedure is used for asymmetric law (pareto law, lognormal law, etc). The results are given in the following table~\ref{tab:one_sample_results1} and table ~\ref{tab:one_sample_results2}

\begin{table}[h!]
	\centering
	\caption{Summary statistics for Normal and Student distributions}
	\label{tab:one_sample_results1}
	\begin{tabular}{|c|c|c|c|c|c|}
		\hline
		Distribution & Trimming & Mean & Confidence & Variance & Confidence \\
		& Level & Estimate &Interval  & Estimate& Interval  \\
		\hline
		\multirow{4}{*}{Normal$(3,2)$}
		& 0.00 & 2.98 & [2.94, 3.02] & 4.01& [3.90, 4.12]  \\
		& 0.05 & 2.98 & [2.95, 3.01] & 2.25& [2.45, 2.56]   \\
		& 0.10 & 2.98 & [2.95, 3.01] & 1.75& [1.71, 1.79]   \\
		& 0.20 & 2.97  &[2.95, 3.00] & 0.85& [0.83, 0.87]   \\
		\hline
		\multirow{4}{*}{5+Student$(df=1)$}
		& 0.00 & 7.16 & [2.79, 11.53] &99606.62& [-91559.29, 290772.53]  \\
		& 0.05 & 4.97 & [4.95, 5.00] & 3.54&  [3.45, 3.64]   \\
		& 0.10 & 4.98 & [4.96, 5.00] & 1.51&  [1.47, 1.54]   \\
		& 0.20 & 4.98  &[4.97, 5.00] & 0.48&  [0.47, 0.49]   \\
		\hline
		\multirow{4}{*}{5+Student$(df=2)$}
		& 0.00 & 4.98 & [4.94, 5.03] & 10.5&  [8.11,12.90]  \\
		& 0.05 & 4.99 & [4.98, 5.01] & 1.29&  [1.26, 1.32]   \\
		& 0.10 & 4.99 & [4.98, 5.01] & 0.76&  [0.74, 0.77]   \\
		& 0.20 & 4.99  &[4.98, 5.00] & 0.32&  [0.31, 0.32]   \\
		\hline
		\multirow{4}{*}{5+Student$(df=3)$}
		& 0.00 & 4.99 & [4.97, 5.02] &2.91&  [2.61, 3.21]  \\
		& 0.05 & 4.99 & [4.98, 5.01] &0.97&  [0.95, 0.99]  \\
		& 0.10 & 4.99 & [4.98, 5.01] &0.61&  [0.60, 0.62]   \\
		& 0.20 & 4.99 & [4.98, 5.00] &0.26&  [0.26, 0.27]   \\
		\hline
		\multirow{4}{*}{5+Student$(df=5)$}
		& 0.00 & 4.99 & [4.98, 5.01] & 1.62& [1.57, 1.68]  \\
		& 0.05 & 4.99 & [4.98, 5.01] & 0.79& [0.78, 0.80]   \\
		& 0.10 & 4.99 & [4.98, 5.00] & 0.52& [0.51, 0.53]   \\
		& 0.20 & 4.99 & [4.98, 5.00] & 0.24& [0.24, 0.25]   \\
		\hline
	\end{tabular}
\end{table}

\begin{table}[h!]
	\centering
	\caption{Summary statistics for Pareto and Lognormal distributions}
	\label{tab:one_sample_results2}
	\begin{tabular}{|c|c|c|c|c|c|}
		\hline
		Distribution & Trimming & Mean & Confidence & Variance & Confidence \\
		& Level & Estimate &Interval  & Estimate& Interval  \\
		\hline
		\multirow{4}{*}{Pareto$(1,1.5)$}
		& 0.00 & 3.00 & [2.82, 3.18] &167.76& [20.89, 314.62]  \\
		& 0.05 & 2.06 & [2.04, 2.07] &1.45 & [1.40, 1.50]   \\
		& 0.10 & 1.89 & [1.88, 1.90] & 0.67& [0.65, 0.68]   \\
		& 0.20 & 1.73  &[1.72, 1.73] & 0.22& [0.21, 0.2]   \\
		\hline
		\multirow{4}{*}{Pareto$(1,2.5)$}
		& 0.00 & 1.67 & [1.65, 1.69] &1.93& [1.33, 2.53]  \\
		& 0.05 & 1.49 & [1.49, 1.50] &0.23& [0.13, 0.14]   \\
		& 0.10 & 1.44 & [1.43, 1.44]& 0.13& [0.12, 0.13]   \\
		& 0.20 & 1.38  &[1.37, 1.38] & 0.05& [0.05, 0.05]   \\
		\hline
		\multirow{4}{*}{Pareto$(1,3)$}
		& 0.00 & 1.50 & [1.49, 1.51] & 0.74& [0.58, 0.89]  \\
		& 0.05 & 1.39 & [1.38, 1.39] & 0.13& [0.13, 0.14]   \\
		& 0.10 & 1.35 & [1.34, 1.35] & 0.07& [0.07, 0.08]   \\
		& 0.20 & 1.30  &[1.30, 1.31] & 0.03& [0.03, 0.03]   \\
		\hline
		\multirow{4}{*}{Lognormal$(0,1)$}
		& 0.00 & 1.64 & [1.61, 1.67] &4.65& [4.17,5.13]  \\
		& 0.05 & 1.28 & [1.26, 1.29] & 1.15& [1.12, 1.18]   \\
		& 0.10 & 1.11 & [1.10, 1.12]& 0.68&  [0.66, 0.69]   \\
		& 0.20 & 0.89  &[0.89, 0.90]  & 0.32& [0.32, 0.33]   \\
		\hline
		\multirow{4}{*}{Lognormal$(3,2)$}
		& 0.00 & 147.63 & [136.42, 158.85] &654994.15& [361274.95, 948713.36]  \\
		& 0.05 & 55.85 & [54.55, 57.16] &8445.13& [8071.17, 8819.09]   \\
		& 0.10 & 38.43 & [37.66, 39.20]&2789.05&  [2693.10, 2885.01]   \\
		& 0.20 & 22.55&[22.16, 22.95] &654.95& [636.41, 673.49]   \\
		\hline
	\end{tabular}
\end{table}

\Ni The results in Table ~\ref{tab:one_sample_results1} and Table ~\ref{tab:one_sample_results2} illustrate the effect of trimming on the estimation of measures of central tendency and dispersion across across different distributions. \\

\Ni For light-tailed distributions like the Normal$(3,2)$, trimming has a negligible influence on the estimated mean, which remains close to the theoretical value of $\mu = 3$. However, the variance decreases with increasing levels of trimming, indicating a slight loss of efficiency from the removal of central observations. Thus, trimming is unnecessary for light-tailed data.\\

\Ni For distributions with moderately heavy tails, such as the Student with low degrees of freedom, trimming dramatically improves the stability and accuracy of both estimators. Particularly, the standard men and variance are highly unstable for Student law with $df=1$, whereas trimmed estimators recover values close to the true mean and yield realistic confidence intervals. The marked reduction in dispersion and confidence interval widths confirms that the Trimmed Functional Empirical Process (TFEP) effectively mitigates the influence of extreme values.\\

\Ni The value of the TFEP becomes particularly evident for skewed and heavily tailed distributions, such as the Pareto and Lognormal distributions. In these cases, untrimmed estimates of the mean and variance are significantly inflated by extreme values. Moderate trimming (5\%–10\%) is sufficient to considerably reduce bias and dispersion, while higher trimming produce robust and stable estimates with narrower confidence intervals.\\

\Ni In summary, these results confirm that trimming within the FEP framework offers an effective compromise between robustness and efficiency. While it provides limited benefit for lightly tailed distributions, it becomes essential for skewed or heavily tailed data, where classical estimators fail to accurately describe central tendency and variability. Thus, the TFEP provides a general and adaptive estimation framework, paving the way for the asymptotic developments.  presented in the following section. \\

\subsection*{Simulation Study: Two-Sample Problem}

\Ni Here we consider some scenarioas composed of two randm independent samples.  

\[
X_1, \dots, X_{n_1} \sim F_1, 
\qquad 
Y_1, \dots, Y_{n_2} \sim F_2,
\]

\Ni with possibly heavy-tailed distributions.  \\

\Ni The following cases are investigated: \\

\begin{itemize}
	\item $X \sim F_1 = \text{Pareto}(1,\alpha_1)$, $Y \sim F_2 = \text{Pareto}(1, \alpha_2)$;\\
	
	\item $X \sim F_1 = \text{Student}(df_1)$, $Y \sim F_2 = \text{Student}(df_2)$;\\
	
	\item $X \sim F_1 = \text{Lognormal}(\mu_1,\sigma_1)$, $Y \sim F_2 = \text{Lognormal}(\mu_2,\sigma_2)$;\\
	
	\item $X \sim F_1 = \text{Normal}(\mu_1=3,\sigma_1=2)$, $Y \sim F_2 = \text{Normal}(\mu_2=0,\sigma_2=1)$ (benchmark).
\end{itemize}

\Bin Sample sizes are chosen as $n_1 = n_2 = 10000$ for each setting. Trimming levels are set to $\tau \in \{0.00, 0.02, 0.05, 0.10, 0.2\}$, ensuring the existence of the trimmed second moments. 

\begin{table}[h!]
	\centering
	\caption{Summary of the statistics, ratio $R_\tau$, mean difference $\Delta \mu_\tau$}
	\label{tab:two_sample_results1}
	\resizebox{\textwidth}{!}{%
		\begin{tabular}{|c|c|c|c|c|}
			\hline
			\multicolumn{5}{|c|}{Normal$(3,2)$ -- Normal$(0,1)$} \\ 
			\hline
			Level. & $R_\tau$ & Confidence Interval & $\Delta\mu_\tau$ & Confidence Interval \\
			\hline
			0.00 & 3.903 & [3.718, 4.097] & 3.002  & [2.958 , 3.045]  \\
			0.05 & 3.89  & [3.725, 4.063] & 2.999  & [2.963, 3.035]  \\
			0.10 & 3.951 & [3.780, 4.131] & 2.995  & [2.963, 3.027]   \\
			0.20 & 4.062 & [3.865, 4.268] & 2.990  & [2.964, 3.016]  \\
			\hline
			\multicolumn{5}{|c|}{Student$(df=1)$ - Student$(df=2)$} \\ 
			\hline
			Level. & $R_\tau$ & Confidence Interval & $\Delta\mu_\tau$ & Confidence Interval \\
			\hline
			0.00 & 1920.279  & [396.920, 9290.205]   & 1.610  & [-1.638, 4.858]  \\
			0.05 & 2.900 & [2.740, 3.069] & 3.011  & [2.960, 3.061]  \\
			0.10 & 2.073 & [1.970, 2.181] & 0.036  & [0.003, 0.070]   \\
			0.20 & 1.537 & [1.460, 1.619] & 0.045  & [0.022, 0.067]  \\
			\hline
			\multicolumn{5}{|c|}{Student$(df=3)$ -- Student$(df=5)$} \\ 
			\hline
			Level. & $R_\tau$ & Confidence Interval & $\Delta\mu_\tau$ & Confidence Interval \\
			\hline
			0.00 & 1.8  & [1.618, 2.003]   & 0.020  & [-0.022, 0.062]  \\
			0.05 & 1.281 & [1.224, 1.341] & 0.021  & [-0.006, 0.049]  \\
			0.10 & 1.250 & [1.194, 1.309] & 0.021  & [-0.003, 0.045]   \\
			0.20 & 1.224 & [1.164, 1.287] & 0.026  & [0.007, 0.044]  \\
			\hline
	\end{tabular}}
\end{table}

\begin{table}[h!]
	\centering
	\caption{Summary of the bootstrap statistics, ratio $R_\tau$, mean difference $\Delta \mu_\tau$}
	\label{tab:two_sample_results2}
	\resizebox{\textwidth}{!}{%
		\begin{tabular}{|c|c|c|c|c|}
			\hline
			\multicolumn{5}{|c|}{Pareto$(1,1.5)$ -- Pareto$(1,2)$} \\ 
			\hline
			Level. & $R_\tau$ & Confidence Interval & $\Delta\mu_\tau$ & Confidence Interval \\
			\hline
			0.00 & 27.239 & [11.225, 66.098] & 0.927  & [0.718, 1.136]  \\
			0.05 & 2.89  & [2.589, 3.004] & 0.368  & [0.340, 0.397]  \\
			0.10 & 2.6   & [2.445, 2.766] & 0.296  & [0.275, 0.316 ]   \\
			0.20 & 2.426 & [2.291, 2.570] & 0.230  & [0.216, 0.243]  \\
			\hline
			\multicolumn{5}{|c|}{Pareto$(1,2.5)$ - Pareto$(1,3)$} \\ 
			\hline
			Level. & $R_\tau$ & Confidence Interval & $\Delta\mu_\tau$ & Confidence Interval \\
			\hline
			0.00 & 2.069  & [1.373, 3.118]   & 0.147  & [0.117, 0.177]  \\
			0.05 & 1.654 & [1.553, 1.763] & 0.098  & [0.085, 0.110 ]  \\
			0.10 & 1.623 & [1.534, 1.717] & 0.084  & [0.075, 0.094]   \\
			0.20 & 1.526 & [1.445, 1.612] & 0.069  & [0.062, 0.076 ]  \\
			\hline
			\multicolumn{5}{|c|}{Lognormal$(1,2)$ -- Lognormal$(1,1)$} \\ 
			\hline
			Level. & $R_\tau$ & Confidence Interval & $\Delta\mu_\tau$ & Confidence Interval \\
			\hline
			0.00 & 340.490  & [191.658, 604.896]   & 15.037  & [12.987, 17.088]  \\
			0.05 & 17.023 & [15.769, 18.378 ] & 4.153  & [3.896, 4.410]  \\
			0.10 & 10.734 & [10.048, 11.467] & 2.459  & [2.293, 2.625]   \\
			0.20 & 6.568 & [6.178, 6.983] & 1.007  & [ 0.911, 1.102]  \\
			\hline
	\end{tabular}}
\end{table}

\Ni Table~\ref{tab:two_sample_results1} and ~\ref{tab:two_sample_results2} present the statistics for the variance ratio $R_\tau$ and the mean difference $\Delta \mu_\tau$, obtained with different levels of trimming. \\

\Ni For normal distributions $\mathcal{N}(3,2)$ and $\mathcal{N}(0,1)$, the estimates are stable and close to their theoretical values, indicating a variance ratio approximatively $4$ and a mean difference around $3$, regardless of trimming level. Similarly, for the Student distributions with with moderate degrees of freedom $df = 3$ and $df = 5$, the both distributions are similar and are comparable, yielding small mean differences and variance ratios close to one; trimming has therefore only a limited influence in these cases.\\

\Ni In contrast, for Student distributions with low degrees of freedom ($df = 1$ versus $df = 2$), for heavy-tailed Pareto and Lognormal distributions, the presence of extreme values severely distorts the untrimmed estimates ($\tau = 0$), leading to very large values of $R_\tau$, $\Delta \mu_\tau$ and wide confidence intervals. The influence of extreme values is substantially reduces when  the trimming level increase. Thus the estimations becomes more robust and stable with narrower confidence intervals. \\

\Ni Overall, these results clearly demonstrate that trimming provides an effective and reliable strategy for improving the robustness of statistical inference when dealing with heavy-tailed or highly skewed data.

\subsection{Empirical Application: Income Data Analysis}

\Ni We apply the Trimmed Functional Empirical Process (TFEP) framework to analyze income data from the ESAM1 (1996) and ESAM2 (2000) surveys for the Dakar and Diourbel regions of Senegal. The datasets, denoted as \texttt{dakar1}, \texttt{dakar2}, \texttt{diour1}, and \texttt{diour2}, provide detailed household income information that exhibits the heavy-tailed characteristics typical of economic data. The characteristics typical of economic data are given in the Table \ref{tab:summary_stats}

\begin{table}[h]
	\centering
	\caption{Summary statistics for income data}
	\begin{tabular}{|lccccccc|}
		\hline
		Dataset & $n$ & Mean & Median & Std.\ Dev. & Skewness & Kurtosis & JB\_p\_value \\
		\hline
		\texttt{dakar1} & 1122 & 500963.5 & 315150.1 & 695638.4 & 6.61 & 61.91& 0\\
		\texttt{dakar2} & 2028 & 673775.9 & 449522.0 & 869884.7 & 7.33 & 79.50 & 0\\
		\texttt{diour1} & 231 & 219368.8 & 164220.1 & 230356.4 & 6.15 & 54.98 & 0\\
		\texttt{diour2} & 573 & 267311.4 & 217093.5 & 183529.7 & 2.86 & 15.67 & 0\\
		\hline
	\end{tabular}
	\label{tab:summary_stats}
\end{table}

\Ni The main descriptive statistics for the income data from Dakar and Diourbel, collected in $1996$ and $2000$ are presented in the Table~\ref{tab:summary_stats}. The statistics concern the measures of central tendency (mean and median), dispersion and shape (skewness, kurtosis), along with the p-value of the Jarque–Bera normality test. \\

\Ni We observe that the incomes levels in Dakar are substantially higher than those in Diourbel. This indicates that economic disparity between the capital and the interior. All datasets present means  notably larger than their corresponding medians, suggesting a strong right-skewed distribution, typical for income data. 

\Ni The standard deviations are also large relative to the means, indicating substantial income dispersion within each region and year. \\

\Ni Regarding the shape parameters, all datasets show positive skewness (ranging from 2.86 to 7.33) and excess kurtosis far greater than the Gaussian benchmark of 3 (ranging from 15.67 to 79.50), indicating the presence of heavy-tailed, leptokurtic distributions. \\

\Ni Finally, the p-values of the the Jarque-Bera test for normality are all zero which provides strong statistical evidence against the null hypothesis. Hence, income distributions in all samples deviate significantly from the Gaussian assumption. This justifies the use of robust or non-Gaussian modeling approaches, such as lognormal, Pareto, or more generally Functional Empirical Process (FEP)–based methods, which are better suited for handling asymmetric and heavy-tailed data.\\

\begin{figure}[h]
	\centering
	\includegraphics[width=0.95\textwidth]{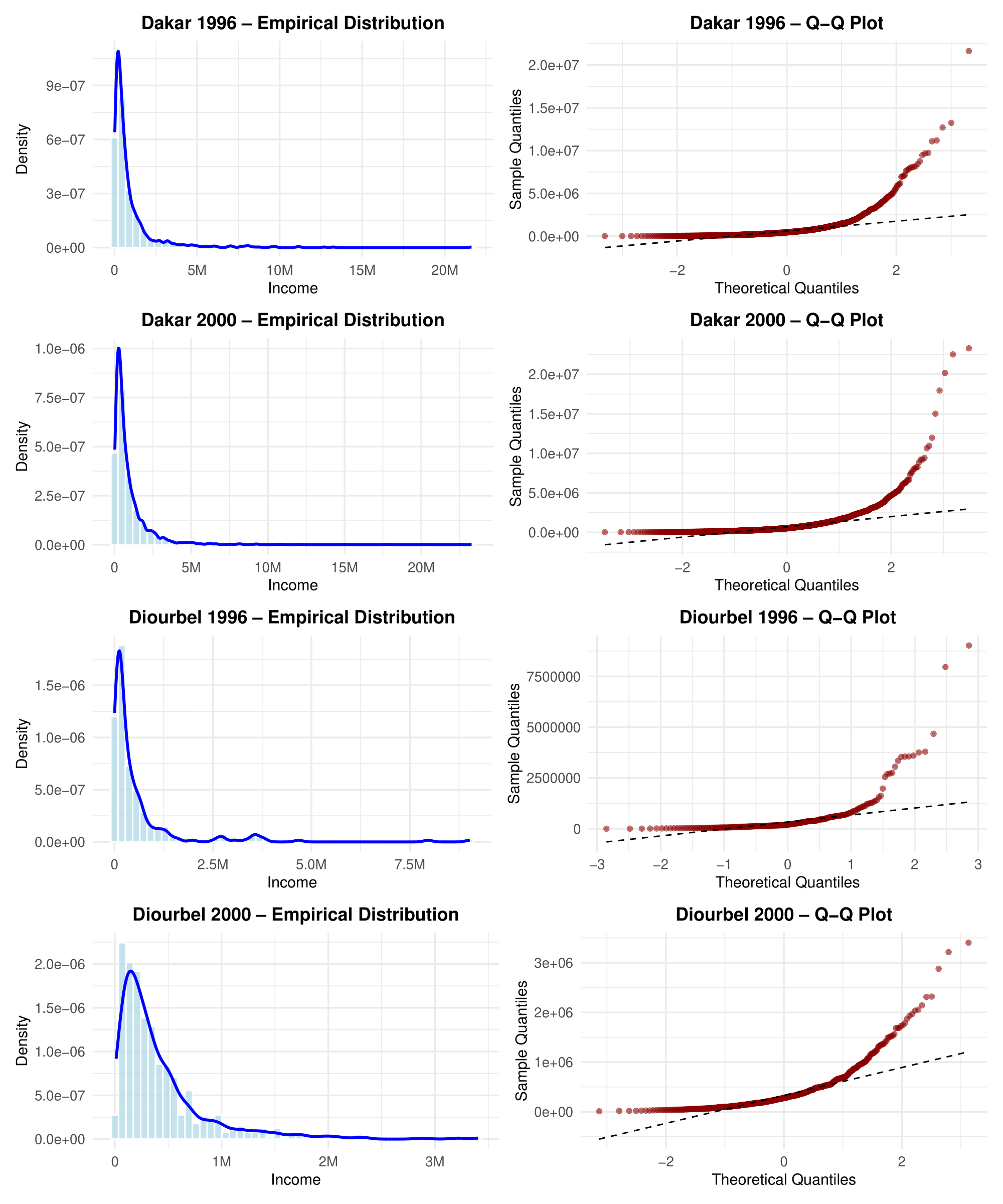}
	\caption{Empirical distributions and Q-Q plots for income data}
	\label{fig:income_dist}
\end{figure}

\Bin Figure \ref{fig:income_dist} in page \pageref{fig:income_dist} clearly demonstrates the non-Gaussian nature of the income data. The Q-Q plots show systematic deviations from the normal distribution, particularly in the upper tails, confirming the need for robust methods like TFEP. \\

\Ni In our study, we apply the methods to randomly selected samples of size $T=200$ from each datasets for analysis. The Jarque-Bera test yields p-values close to zero for all samples, indicating strong departures from normality. Furthermore, the kurtosis values exceed $K > 16$, revealing extremely leptokurtic distributions.

\Ni Now, we formulate the following research questions as statistical hypotheses:

\begin{enumerate}
	\item \textbf{Regional comparison}: $H_0: \mu_{\text{Dakar}} = \mu_{\text{Diourbel}}$ vs $H_1: \mu_{\text{Dakar}} \neq \mu_{\text{Diourbel}}$ for each survey year. \\
	
	\item \textbf{Temporal evolution}: $H_0: \mu_{1996} = \mu_{2000}$ vs $H_1: \mu_{1996} \neq \mu_{2000}$ for each region.
\end{enumerate}

\Ni We first treat the TFEP estimation of the mean income for Dakar and Diourbel in the two distincts periods $1996$ and $2000$. The results are given in the Table \ref{tab:mean_estimates}, with Width1  confidence interval for classical FEP $(0\%)$ and Width2 confidence interval for Trimming FEP $(10\%)$. 

\begin{table}[h]
	\centering
	\caption{TFEP estimates of mean income (in thousands of CFA francs) with 95\% confidence intervals}
	\begin{tabular}{|c|c|c|c|c|}
		\hline
		Dataset & FEP[CI] & TFEP ($10\%$)[CI] & Width1 & Width2 \\
		\hline
		
		\hline
		\texttt{dakar1} & 458.057 [357.085, 559.027] & 322.646 [299.514, 345.778] &201.942 &46.264 \\
		\texttt{dakar2} &688.809[524.271, 853.347] &454.699[422.451, 486.947] &329.076 & 64.496 \\
		\texttt{diour1} & 214.216 [189.094, 239.338 ] & 165.791[155.113, 176.469] &50.244&21.356 \\
		\texttt{diour2} &261.962[234.387, 289.536] & 213.241[199.650, 226.832] &55.149&27.182 \\
		\hline
	\end{tabular}
	\label{tab:mean_estimates}
\end{table}

\Bin The results in the Table \ref{tab:mean_estimates}  highlight the efficiency gains of TFEP over classical methods. The TFEP confidence intervals are very narrower than classical intervals, reflecting the variance reduction achieved through trimming.  Thus the TFEP approach effectively balances robustness and precision, providing more reliable mean estimates for non-Gaussian, heavy-tailed data such as income distributions. This justifies use of TFEP approach as a robust alternative to the classical FEP. \\

\Ni Next, we treat the regional income comparison using the difference in mean and variance ratio. The results are shown in Table~\ref{tab:ci_mu} and Table~\ref{tab:ci_sigma}.\\

\begin{table}[h!]
	\centering
	\caption{Estimation and Confidence Interval TFEP for Difference means of Variances using TFEP}
	\label{tab:ci_mu}
	\begin{tabular}{|c|c|c|c|}
		\hline
		Methods & $\Delta\mu_\tau$ & Confidence Interval& Width \\
		\hline
		\multicolumn{3}{|c|}{Dakar(1996 vs 2000)}& \\ 
		\hline
		Classical FEP (0\%) &230752.562 &[37703.411,423801.713]& 386098.3\\
		Trimming FEP (10\%) &132053.698 & [92367.268,171740.129]& 79372.86\\
		\hline
		\multicolumn{3}{|c|}{Dakar vs Diourbel (1996)}& \\ 
		\hline
		Classical FEP (0\%)&243840.543&[139790.896,347890.190]&  208099.3\\
		Trimming FEP (10\%) &156854.747& [131377.214,182332.281]& 50955.07\\
		\hline
		\multicolumn{3}{|c|}{Dakar vs Diourbel (2000)}& \\ 
		\hline
		Classical FEP (0\%)&426847.535 &[260014.892,593680.178]& 333665.3 \\
		Trimming FEP (10\%) &241458.221&[206463.442,276453.001]& 69989.56\\
		\hline
		\multicolumn{3}{|c|}{Diourbel (1996 vs 2000)}& \\ 
		\hline
		Classical FEP (0\%)&47745.570 &[10442.954,85048.186]&74605.23  \\
		Trimming FEP (10\%) &47450.225 &[30166.517,64733.932]& 34567.42\\
		\hline
	\end{tabular}
\end{table}

\begin{table}[ht]
	\centering
	\caption{Estimation and Confidence Interval TFEP for Ratio of Variances using TFEP}
	\label{tab:ci_sigma}
	\resizebox{\textwidth}{!}{%
		\begin{tabular}{|c|c|c|c|}
			\hline
			Methods & $R_\tau$ & Confidence Interval& Width \\
			\hline
			\multicolumn{4}{|c|}{Dakar(1996 vs 2000)} \\ 
			\hline
			Classical FEP (0\%) & 2.655 & [0.424, 16.612] &16.188\\
			Trimming FEP (10\%) & 1.943 & [1.358, 2.782 ] &1.424 \\
			\hline
			\multicolumn{4}{|c|}{Dakar vs Diourbel (1996)} \\ 
			\hline
			Classical FEP (0\%) & 16.154  & [3.360, 77.665] &74.305\\
			Trimming FEP (10\%) & 4.693   & [3.271, 6.733]  &3.462 \\
			\hline
			\multicolumn{4}{|c|}{Dakar vs Diourbel (2000)} \\ 
			\hline
			\hline
			Classical FEP (0\%) & 35.605  & [9.485, 133.649] &124.164\\
			Trimming FEP (10\%) & 5.630   & [3.920, 8.087]  &4.167 \\
			\hline
			\multicolumn{4}{|c|}{Diourbel (1996 vs 2000)} \\ 
			\hline
			Classical FEP (0\%) & 1.205  & [0.478, 3.035] &2.557\\
			Trimming FEP (10\%) & 1.620   & [1.125, 2.332]&1.207 \\
			\hline
	\end{tabular}}
\end{table}

\Bin The results in the Table~\ref{tab:ci_mu} show that for each scenario, the confidence intervals for estimated $\Delta\mu$ are more precise with smaller amplitude after trimming process. Thus, the Trimmed Functional Empirical Process constitutes a robust alternative for estimating differences in means in socio-economic contexts characterized by skewed or heavy-tailed distributions. In contrast, the results in the Table~\ref{tab:ci_sigma} demonstrate that the TFEP approach significantly improves the reliability of statistical inference on variance ratios in contexts of skewed or thick-tailed distributions, by limiting the influence of outliers.\\

\begin{figure}[h]
	\centering
	\includegraphics[width=0.95\textwidth]{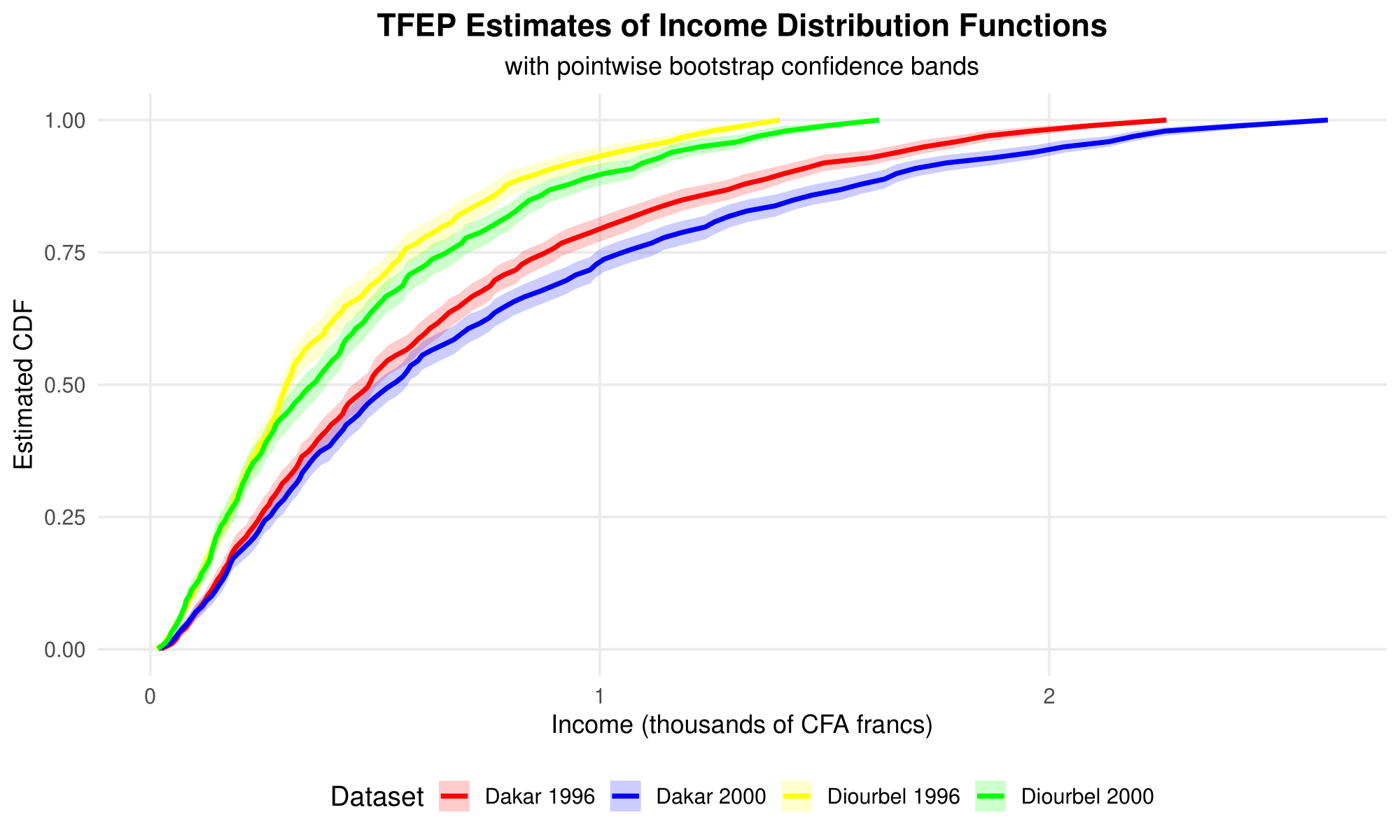}
	\caption{TFEP estimates of income distribution functions with pointwise confidence bands}
	\label{fig:cdf_comparison}
\end{figure}

\Ni According to the results in the Table~\ref{tab:ci_mu} and  Table~\ref{tab:ci_sigma}, Dakar incomes are significantly higher than Diourbel in both survey years, so Dakar distributions dominate Diourbel distributions across most of the income range see the Figutre~\ref{fig:cdf_comparison} in page \pageref{fig:cdf_comparison}. That means incomes are structurally higher in Dakar than in Diourbel. Both regions experienced statistically significant income growth between 1996 and 2000 (temporal growth) and the income gap between Dakar and Diourbel remains substantial across both periods. In Diourbel, the income distribution remained stable between the two period 1996 and 2000.\\

\newpage
\section{Conclusion}\label{sec_06}

\Ni This work  introduces the Trimmed Functional Empirical Process (TFEP) as a robust framework for statistical inference with heavy-tailed distributions. By trimming extreme order statistics, the TFEP regularizes the empirical process, enabling weak convergence to a Gaussian limit even when the population variance is infinite, a case where the standard Functional Empirical Process fails. We establish the weak convergence of TFEP under general regularity conditions and derive asymptotically normal statistics for trimmed means, variances, mean differences, and variance ratios. These results provide a unified basis for one- and two-sample inference with non-standard data. Monte Carlo experiments is used to estimate the difference in  means and variance ratio, the confidence intervals for the difference in means and variance ratio are more precise when the trimming levels increase.  We say that the trimming stabilizes statistical inference under heavy-tailed and skewed distributions. \\

\Ni Application to Senegalese real income data confirms these findings: TFEP reduces bias from extreme values and yields precise estimates of mean differences and variance ratios. The analysis reveals a significant increase in average income in Dakar between 1996 and 2000, highlighting the role of trimming in uncovering structural patterns and enabling reliable group comparisons. \\

\Ni In summary, TFEP provides a rigorous and practical methodology for robust moment inference in non-Gaussian settings, balancing outlier mitigation with sample representativeness.

\end{document}